\documentclass[%
 reprint,
superscriptaddress,
nobibnotes,
 amsmath,amssymb,
 aps,
prl,
showkeys
]{revtex4-2}
\usepackage{graphicx}  
\usepackage{dcolumn}   
\usepackage{bm}        
\usepackage{amssymb}   
\usepackage{siunitx}
\usepackage[breaklinks=true,colorlinks,citecolor=blue,linkcolor=red,urlcolor=blue]{hyperref}
\usepackage{multirow}
\usepackage{xcolor}
\usepackage{amsmath}
\usepackage{hhline}
\usepackage{lipsum}

\usepackage{soul}
\usepackage{color} 

\usepackage{braket}
\usepackage{autobreak}

\usepackage{amsfonts}
\usepackage{amsmath}
\usepackage{slashed}
\usepackage{amsthm}
\usepackage{amssymb}

\definecolor{olive}{rgb}{0.0, 0.6, 0.0}

\definecolor{mycolor}{rgb}{0.0, 0.0, 1.0}
 
\begin{document}

\title{Engineering Majorana Kramers Pairs In Synthetic High Spin Chern Insulators}


\author{Yi-Chun Hung}
\email[Contact author: ]{hung.yi@northeastern.edu}
\affiliation{Department of Physics, Northeastern University, Boston, Massachusetts 02115, USA}
\affiliation{Quantum Materials and Sensing Institute, Northeastern University, Burlington, Massachusetts 01803, USA}

\author{Chen-Hsuan Hsu}
\affiliation{Institute of Physics, Academia Sinica, Taipei 115201, Taiwan}

\author{Arun Bansil}
\email[Contact author: ]{ar.bansil@northeastern.edu}
\affiliation{Department of Physics, Northeastern University, Boston, Massachusetts 02115, USA}
\affiliation{Quantum Materials and Sensing Institute, Northeastern University, Burlington, Massachusetts 01803, USA}


\begin{abstract}
{High spin-Chern-number topological phases provide a promising low-dimensional platform for realizing double-helical edge states. 
In this letter, we show how these edge states can host a variety of phases driven by electron interaction effects, including multi-channel helical Luttinger liquid, spin density wave, superconducting phases, and a new type of $\pi$-junction analog of the latter two, where the transitions between the phases can be controlled. The superconducting phase in the interacting system is shown to be adiabatically connected to a time-reversal-symmetric topological superconductor in the non-interacting DIII class. This connection stabilizes Majorana Kramers pairs as domain wall states at the interface between the superconducting and $\pi$-spin-density wave phases, with the latter exhibiting a time-reversal-symmetric spin-density wave phase.
We discuss the possibility of realizing our proposed scheme for generating Majorana Kramers pairs in a cold-atom based platform with existing techniques, and how it could offer potential advantages over other approaches.
}

\end{abstract}

\maketitle

\par
\begin{figure}[t]
  \centering
  \centering
    \includegraphics[width=\linewidth]{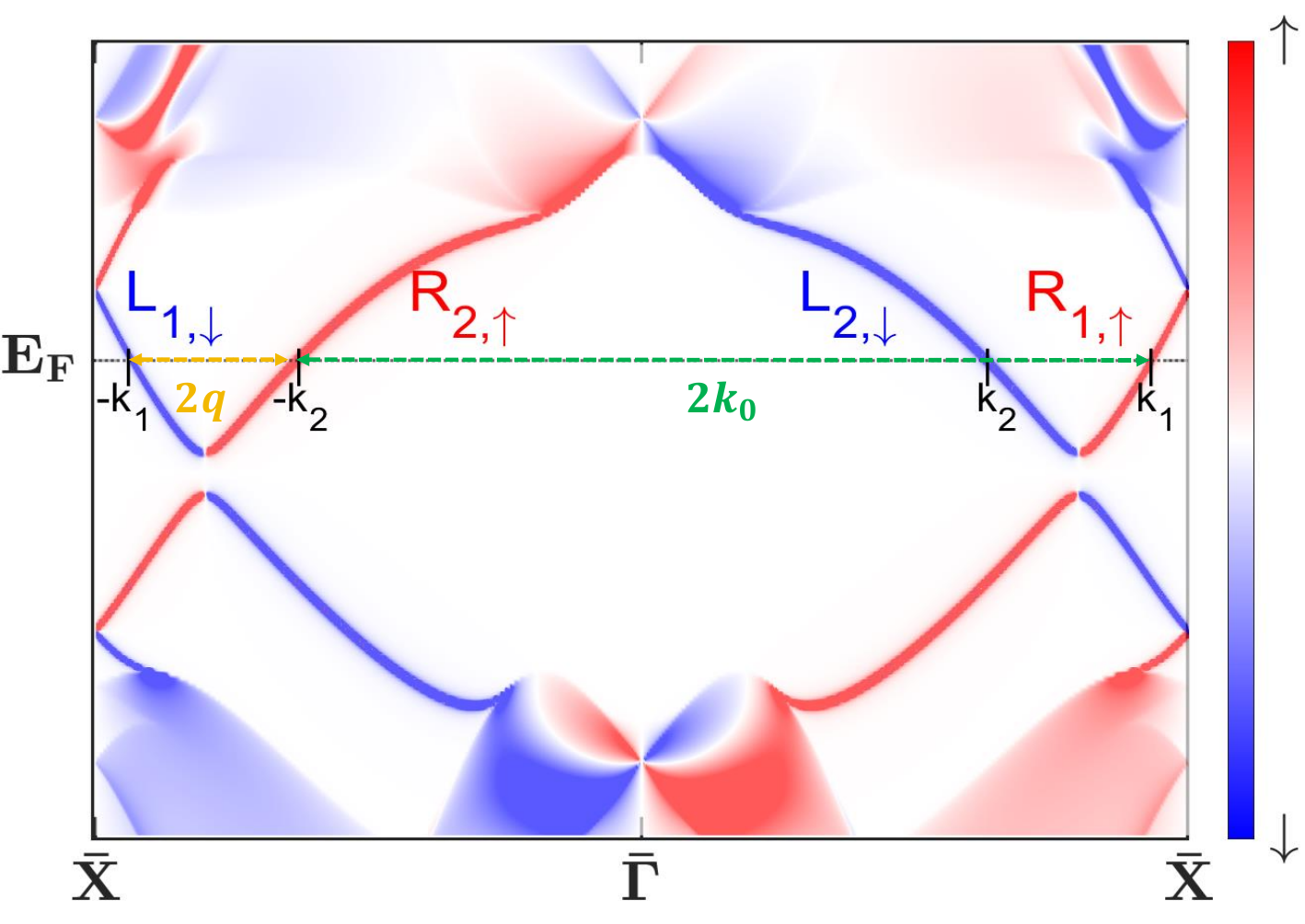}
  \caption{Spin-resolved energy spectrum of the extended Haldane model \cite{PhysRevB.110.035125, PhysRevB.104.L201110} hosting SCI phase with $|C_s|=2$. Two pairs of helical edge states are seen to appear at the Fermi level, $\text{E}_\text{F}$ (measured from the edge Dirac point). The right-(left)-moving modes are labeled as $R_{\mu,\uparrow}$ ($L_{\mu,\downarrow}$), where $\mu=1,2$ denotes the time-reversal sector. Yellow and green arrows indicate the nesting vector $2q=k_1-k_2$ of the SDW and the transferred momentum $2k_0=k_1+k_2$ of the associated forward-scattering process, respectively.  
  }
  \label{fig:01}
\end{figure}

 Recent research on spin Chern insulators (SCIs) 
\cite{PhysRevB.109.155143, Wang_2024, PhysRevB.110.L161104, PhysRevB.110.035161, PhysRevB.94.235111, Kang:2024a, Kang:2024b, campos2024dual} and mirror Chern insulators (MCIs) \cite{Dziawa2012, Tanaka2012, Liu2014, Liu2015, campos2024dual} has underscored their potential for realizing topological phases characterized by higher even spin or mirror Chern numbers, $C_s$ or $C_M \in 2\mathbb{Z}$. 
SCIs and MCIs thus offer a platform for realizing double helical states  in which two pairs of time-reversal-symmetric counter-flowing modes appear within a single edge (FIG.~\ref{fig:01}). 
These double helical states originate from the non-trivial ground-state topology \cite{PhysRevB.80.125327, shulman2010robust, wang2023feature, Lin2024, PhysRevLett.106.106802} and can support higher-order topological phases \cite{PhysRevB.108.245103, PhysRevB.110.035125, PhysRevB.104.L201110, PhysRevB.110.035161}. Promising realizations of SCIs with $|C_s|=2$ include bilayer $\text{Bi}_\text{4}\text{Br}_\text{4}$ \cite{PhysRevB.109.155143}, $\alpha$-antimonene \cite{Wang_2024}, $\text{Ru}\text{Br}_\text{3}$ \cite{PhysRevB.110.L161104, PhysRevB.110.035161}, $(\text{M}\text{O}_\text{2})_\text{2}(\text{Zr}\text{O}_\text{2})_\text{4}$ \cite{PhysRevB.94.235111} with $\text{M}=\text{Pt},\text{W}$, 
twisted bilayer MoTe$_2$ \cite{Kang:2024a} and WSe$_{2}$ \cite{Kang:2024b}, and $\text{Na}_\text{2}\text{Cd}\text{Sn}$ \cite{campos2024dual}. Additionally, $\text{Na}_2\text{Cd}\text{Sn}$ \cite{campos2024dual} and thin films of the $\text{Sn}\text{Te}$ family \cite{Dziawa2012, Tanaka2012, Liu2014, Liu2015} have been demonstrated to be MCIs with $|C_M|=2$. Although the non-trivial topology in these candidate materials has been established, the effects of interactions on their helical edge states have remained unexplored.

\par Helical edge states can stabilize Majorana zero modes (MZMs) through the superconducting proximity effect, which can be triggered by introducing time-reversal-symmetry (TRS) breaking perturbations, such as magnetic impurities or external magnetic fields \cite{PhysRevLett.123.036802, PhysRevLett.124.227001, PhysRevLett.123.167001, PhysRevB.98.245413, PhysRevResearch.2.013330, PhysRevLett.125.017001, PhysRevB.108.184505, PhysRevB.97.205134, yin2024multifoldmajoranacornermodes, PhysRevB.99.020508}. However, magnetism is typically detrimental to superconductivity.
In contrast, Majorana Kramers pairs (MKPs), which are composed of two MZMs linked by TRS, are increasingly drawing attention for their potential in the development of fault-tolerant qubits, crucial for advancing topological quantum computation \cite{PhysRevB.93.045417, PhysRevLett.100.096407, PhysRevLett.120.267002, PhysRevLett.129.227002, PhysRevLett.115.237001, PhysRevB.89.220504, PhysRevLett.112.126402, RevModPhys.80.1083, PhysRevX.4.021018, PhysRevB.106.014522, PhysRevB.94.224509, HAIM20191, PhysRevB.96.035306, 10.1063/5.0102999, PhysRevB.86.184516, PhysRevB.96.081301, PhysRevX.4.021018}. Recent proposals include exploiting helical edge states of (higher-order) topological insulators \cite{PhysRevLett.121.196801, PhysRevB.90.155447, PhysRevB.102.195401, Hsu_2021, PhysRevB.102.205402, PhysRevB.100.165420, PhysRevB.101.104502, PhysRevLett.121.186801, PhysRevLett.121.096803, PhysRevB.106.085420, D4NH00254G}. This approach for realizing MKPs is based on the superconducting proximity effect involving unconventional pairings ($d$-, $p$-, and $s_\pm$-wave) or the presence of a dominant nonlocal pairing mechanism, and presents experimental challenges involving materials and their intricate interface engineering. Furthermore, phonons can impede nonlocal pairings in helical channels \cite{D4NH00254G} and destabilize the topological zero modes. Therefore, it is highly desirable to find schemes that avoid the complications associated with the superconducting proximity effect. 

While realization of MKPs in solid-state systems remains elusive, cold atom systems offer a viable alternative by providing tunable effective interactions~\cite{RevModPhys.82.1225, PhysRevLett.93.050401, PhysRevA.92.053612, Mitra2018, PhysRevA.73.042705, PhysRevLett.91.020402, PhysRevA.85.051602} and interlayer tunneling~\cite{PhysRevLett.125.010403, Gall2021, samland2024thermodynamicsdensityfluctuationsbilayer, Meng2023}.
This has inspired proposals for generating various topological states, such as the helical edge states emerging from the quantum spin Hall effect~\cite{PhysRevLett.109.205303, PhysRevLett.105.255302, PhysRevLett.111.225301, Scheurer2015, PhysRevLett.97.240401, PhysRevLett.109.145301} and MKPs~\cite{PhysRevLett.123.060402, PhysRevA.103.013307, Ye2017}, which led to the experimental realization of quantum spin Hall insulators~\cite{PhysRevLett.111.185301}.
These proposals, however, are also experimentally challenging because they require line solitons across the entire system~\cite{PhysRevLett.123.060402} and the control of spin-orbit coupling~\cite{PhysRevA.103.013307, Ye2017}. 

\par In this letter, we discuss the double helical edge states of SCIs and MCIs, and demonstrate that electron-electron interactions within the edges can give rise to a rich tapestry of exotic phases, including multi-channel helical Luttinger liquids (LL), spin density waves (SDWs), and superconducting (SC),  $\pi$-SDW, and $\pi$-SC phases. Notably, we show that the system in the SC phase can be adiabatically connected to a time-reversal-symmetric topological superconductor in the non-interacting DIII class~\cite{RevModPhys.88.035005, PhysRevB.84.060504,PhysRevB.82.115120}. 
This connection reveals the emergence of MKPs in interacting systems
without relying on the superconducting proximity effect, manifesting as domain wall states at the boundaries between the SC and $\pi$-SDW phases. 
Finally, we discuss how our proposed scheme could be realized experimentally in cold-atom based optical lattices with existing techniques.

\paragraph{Model---} We focus on a relatively simple example of double helical states related by TRS; see FIG.~\ref{fig:01}. The two pairs of counterflowing modes can be represented by the fermion fields as: 
\begin{equation}\label{eq:01}
\psi_\mu(r) = R_{\mu\uparrow}(r)e^{i\xi_\mu k_\mu r} + L_{\mu\downarrow}(r)e^{-i\xi_\mu k_\mu r},
\end{equation}
where the index $\mu$ labels the sector of time-reversal partner, with $\xi_1=1$ and $\xi_2=-1$. We assume that the Fermi level is far from the Dirac point of the edge spectrum such that $|k_1-k_2| L \gg 1 $ with the edge length $L$. Due to the helical nature of the edge states, below we suppress the spin indices for the slow oscillating modes, $R_\mu , L_\mu$. The Hamiltonian of the low-energy states can be expressed as $H=H_0+ H_\text{int}$, with the kinetic energy term,
\begin{equation}\label{eq:02}
    H_0 = -i\hbar v\int dr \,\sum_{\mu=1,2} \big[ R^\dagger_\mu(r)\partial_r R_\mu(r) - L^\dagger_\mu(r)\partial_r L_\mu(r) \big],
\end{equation}
with the Fermi velocity $v$. 
Here, we assume a negligible difference in Fermi velocities \cite{Dziawa2012, Tanaka2012, Liu2014, Liu2015, PhysRevB.109.155143, Wang_2024, PhysRevB.110.L161104, PhysRevB.110.035161, PhysRevB.94.235111} and a uniform chemical potential for the two time-reversal sectors \cite{PhysRevB.110.035125}. 
Moreover, we consider general density-density and current-current interactions between electrons:
\begin{equation}\label{eq:03}
    H_\text{int} = \frac{1}{2}\int drdr' \big[ V(r-r')\rho(r)\rho(r') + \tilde{V}(r-r')J(r)J(r') \big],    
\end{equation}
with the charge ($\rho$) and current ($J$) densities and their corresponding short-range strengths, $V$ and $\tilde{V}$.

\par Following a standard procedure \cite{book:Giamarchi}, we bosonize Eqs.~\eqref{eq:02}--\eqref{eq:03} to obtain a two-channel LL, $H_{\text{LL}}$, and a sine-Gordon-like mass term, $H_{\text{m}}$, 
\begin{align}
    H_{\text{LL}} = & \sum_{\mu=c,\tau} \int \frac{\hbar\, dr }{2\pi} u_\mu[K_\mu^{-1}(\partial_r\varphi_\mu(r))^2 + K_\mu(\partial_r\vartheta_\mu(r))^2],  \label{eq:05}
    \\ 
    H_{\text{m}} = & \int dr\,\frac{m}{2(\pi a)^2}\cos[2\sqrt{2}\vartheta_\tau(r)] \label{eq:06},
\end{align}
with the LL parameters, $K_c,K_\tau$, and the renormalized velocities, $u_c,u_\tau$. 
The indices $c$ and $\tau$ are associated with the charge and pseudospin sectors, respectively. These names reflect the nature of the collective excitations, which maintain helical characteristics within each sector,   see Supplemental Materials (SM) and FIG.~S3 for details \cite{SM}\nocite{PhysRevB.61.13410, PhysRevB.102.165147, PhysRevLett.103.166403, PhysRevB.109.155119, PhysRevB.99.045125, PhysRevB.98.115146, PhysRevB.86.165121, PhysRevB.97.045415, J_M_Kosterlitz_1974, RevModPhys.87.457, annurev_PDW, PhysRevB.79.064515, PhysRevB.91.195102, PhysRevB.101.165133}. 
The mass  $m\equiv V_{2k_0}-\tilde{V}_{2k_0}$ above is determined by the Fourier components, $V_{2k_0}$ and $\tilde{V}_{2k_0}$, of the potential $V$ and $\tilde{V}$, respectively, corresponding to the transferred momentum $\delta k = 2k_0$ in the forward-scattering process depicted in FIG.~\ref{fig:01}. Below, we show how the mass term induces a SDW or an SC phase within the edge. 

\begin{figure}[t]
  \centering
  \centering
    \includegraphics[width=\linewidth, height = 6.5 cm]{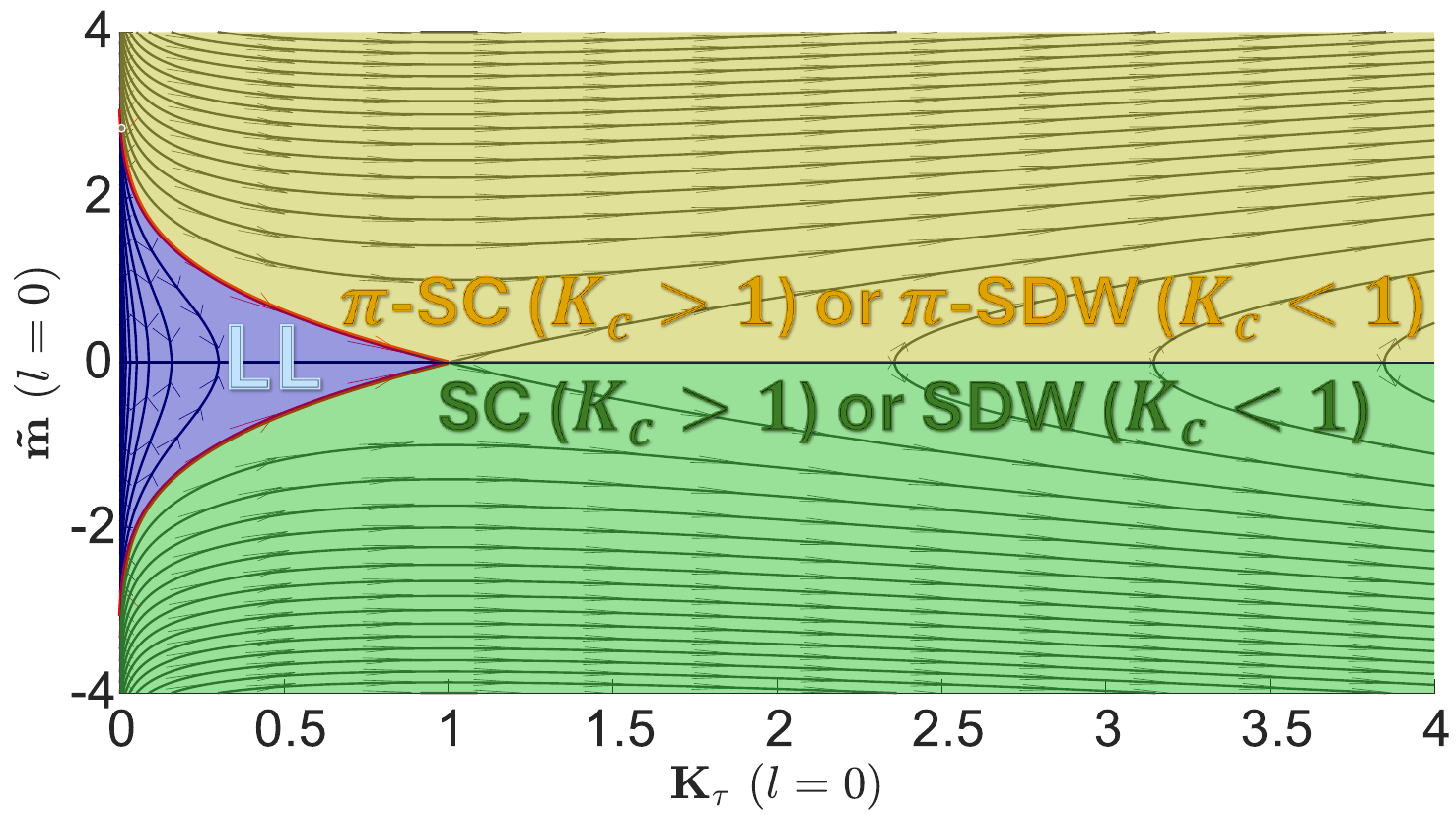}
  \caption{
  Phase diagram.   
Blue, green, and yellow colors indicate the regions of LL, SC ($K_c>1$) or SDW  ($K_c<1$), and $\pi$-SC ($K_c>1$) or $\pi$-SDW ($K_c<1$) phases, respectively. Arrows indicate the direction of RG flows. The red solid line indicates the critical line, $f(\tilde{m}(l=0))$ in Eq.~\eqref{eq:rel_m}, for $\tilde{m}$ to be relevant \cite{SM}.
  }
  \label{fig:03}
\end{figure}

\paragraph{Phase diagram---} We derive the renormalization-group (RG) flow equations and get the following condition for a relevant mass term \cite{SM}, 
 \begin{align}
\label{eq:rel_m}
K_\tau(l=0) > f(\tilde{m}(l=0)) \text{ with } \tilde{m}(l=0)\neq0,
\end{align}
with $\tilde{m}\equiv m/\hbar v$, $l\equiv\ln [ a(l)/a(0) ]$, the cutoff $a$, and 
$f(x)$ satisfying $x^2 = 2 [f(x) - \ln f(x) ]-2$ and $f(x)\leq1$.
The corresponding RG flow is depicted in FIG.~\ref{fig:03}.
Given that in practice $K_\tau(l=0) = 1$~\cite{SM}, the mass term remains relevant as long as there is a finite mass in Eq.~\eqref{eq:06}. Below we will focus on this case.

\par In the strong-coupling limit, a relevant $\tilde{m}$ term in Eq.~\eqref{eq:06} can pin the $\vartheta_\tau $ field at a certain value depending on the sign of $m$. In this limit, we examine the following operators characterizing the SDW and SC instabilities, 
\begin{equation}
\begin{split}
    \Delta_{2q}^{(x)} (r) \pm i\Delta_{2q}^{(y)} (r) 
    & \sim \mp e^{\pm i[\sqrt{2}\varphi_c(r)]}\cos[\sqrt{2}\vartheta_\tau(r)],
    \\ 
    \Delta_{\text{SC}} (r)
    & \sim e^{- i[\sqrt{2}\vartheta_c(r)]}\cos[\sqrt{2}\vartheta_\tau(r)] ,
\end{split}
\end{equation}
respectively. Here, the superscript and subscript of the SDW operators denote the components and their nesting vector (see FIG.~\ref{fig:01}), respectively. Interestingly, the double helical states allow for a generalization of the SDW and SC phases to include a $\pi$ phase difference between the two time-reversal sectors. In analogy to the $\pi$-junction setting~\cite{PhysRevLett.115.237001, Hsu_2021, PhysRevLett.100.096407, PhysRevLett.111.116402, PhysRevLett.122.126402, PhysRevLett.120.267002, PhysRevLett.129.227002, PhysRevB.97.014507, PhysRevB.89.220504}, we refer to these as the $\pi$-SDW and $\pi$-SC orderings, with the following operators, 
\begin{equation}
\begin{split}
    \Delta_{2q,\pi}^{(x)} (r) \pm i\Delta_{2q,\pi}^{(y)} (r) 
    & \sim ie^{\pm i[\sqrt{2}\varphi_c(r)]}\sin[\sqrt{2}\vartheta_\tau(r)],
    \\ 
    \Delta_{{\rm SC},\pi} (r)
    & \sim ie^{- i[\sqrt{2}\vartheta_c(r)]}\sin[\sqrt{2}\vartheta_\tau(r)],
\end{split}
\end{equation}
respectively. Note that $\pi$-SDW respects the TRS, see SM and FIG.~S1 for details \cite{SM}. When \(m<0\), \(\vartheta_\tau \) is pinned at \(n\pi / \sqrt{2}\) with \(n\in\mathbb{Z}\), resulting in \(\langle \cos(\sqrt{2}\vartheta_\tau) \rangle \neq 0\) and \(\langle \sin(\sqrt{2}\vartheta_\tau) \rangle = 0\), suppressing the $\pi$-SDW and $\pi$-SC phases. On the other hand, when \(m>0\), \(\vartheta_\tau \to (n+\frac{1}{2}) \pi / \sqrt{2}\), and the SDW and SC phases are suppressed instead.

\par Since the $\vartheta_\tau$ field is pinned to its classical value, the correlation functions of these operators exhibit power-law decay, modulated by the remaining $\varphi_c$ and $\vartheta_c$ fields, allowing us to investigate the instabilities in the edge. Remarkably, LL becomes unstable towards the SDW (or $\pi$-SDW) and SC (or $\pi$-SC) phases for $K_c < 1$ and $K_c > 1$, respectively, see SM and FIG.~S2 for details \cite{SM}. Under these conditions, either $\varphi_c$ or  $\vartheta_c$ field will also be ordered, further gapping out the charge sector. In mesoscopic systems, this leads to the establishment of quasi-long-range orders~\cite{book:Giamarchi, PhysRevB.92.035139}.

Figure~\ref{fig:03} summarizes our results for the phase diagram. In common solid-state systems, repulsive Coulomb interactions lead to the $\pi$-SDW phases \cite{SM}. One could use cold-atom systems to control both the interaction strength and the sign of $m$, offering an alternative platform for manipulating the phases in the helical edge states, which we discuss later.
  

\paragraph{Emergent Majorana Kramers pair---} 
We turn now to show that MKPs appear as the domain wall states between the SC and $\pi$-SDW phases. Note from FIG.~\ref{fig:03} that the points at $K_\tau=1$ with $m\neq0$ can be adiabatically connected to those at $K_{\tau}=2$ through the RG flows without closing the energy gap \cite{SM}. At the latter points, we can refermionize the pseudospin sector to study the emergent fermionic soliton excitations within this sector~\cite{PhysRevLett.113.256405},
\begin{equation}
\label{eq:23}
\begin{split}
    H_\tau(r) = & -i\hbar v [\mathcal{R}_{\tau}^\dagger(r)\partial_r\mathcal{R}_{\tau}(r) - \mathcal{L}_{\tau}^\dagger(r)\partial_r\mathcal{L}_{\tau}(r)]
    \\ & + \frac{\Delta_m}{2} [\mathcal{R}_{\tau}^\dagger(r)\mathcal{L}_{\tau}^\dagger(r) - \mathcal{L}_{\tau}^\dagger(r)\mathcal{R}_{\tau}^\dagger(r) + \text{H.c.}] , 
\end{split}
\end{equation}
where $\mathcal{R}_{\tau} = \frac{\kappa_R}{\sqrt{2\pi a}}e^{i (\vartheta_\tau' -\varphi_\tau' )}$ and $ \mathcal{L}_{\tau} = \frac{\kappa_L}{\sqrt{2\pi a}}e^{i (\vartheta_\tau'+\varphi_\tau')}$ are the right- and left-moving fields, respectively. We have $\vartheta_\tau' \equiv\sqrt{2}\vartheta_\tau $, $\varphi_\tau'\equiv\varphi_\tau /\sqrt{2}$, and $\Delta_m\equiv\frac{m^*}{2\pi a}$ with renormalized $m^*$. Here, the fields $\mathcal{R}_{\tau}$ and $ \mathcal{L}_{\tau}$ preserve the helical nature of the collective excitations in the pseudospin sector \cite{SM}, and we have suppressed their spin indices for simplicity.

The refermionized expression allows us to further explore its topological properties. Specifically, Eq.~\eqref{eq:23} can be recast into the Bogoliubov-de Gennes (BdG) form in the basis of 
 $( \mathcal{R}_{\tau},  \mathcal{L}_{\tau},  \mathcal{R}_{\tau}^\dagger,  \mathcal{L}_{\tau}^\dagger )^T$,
\begin{equation}
\label{eq:25}
    H_\tau^{(\text{BdG})}(r) = -i\hbar v \eta_0\sigma_3 \partial_r 
    - \Delta_m\eta_2\sigma_2,
\end{equation}
where $\eta_\mu$ and $\sigma_\mu$ are the Pauli matrices for the particle-hole and spin degree of freedom, respectively. It can be shown that the non-interacting fermionic model in Eq.~\eqref{eq:25} preserves spinful TRS (with $\Theta=i\sigma_2 K$), particle-hole symmetry (with $\Xi=\eta_1 K$) and the chiral symmetry (with $\Gamma=\Theta\Xi$). Therefore, it belongs to DIII class in the Altland-Zirnbauer symmetry classification \cite{RevModPhys.88.035005, PhysRevB.84.060504, PhysRevB.82.115120} and it is characterized by a $\mathbb{Z}_2$ topological invariant \cite{SM},
\begin{equation}
\label{eq:26}
    \nu=\frac{1-\text{sign}(\Delta_m)}{2},
\end{equation}
obtained by regularizing Eq.~\eqref{eq:25} into a topologically equivalent lattice model \cite{PhysRevB.82.115120}. Considering a domain wall separating the SC (i.e., $\Delta_m<0$) and $\pi$-SDW (i.e., $\Delta_m>0$) phases within the helical edge, we solve the corresponding BdG equation for Eq.~\eqref{eq:25} and find zero-energy MKP localized at the domain wall~\cite{SM}, consistent with Eq.~\eqref{eq:26}. Note that since the SDW phases spontaneously break TRS, it is necessary to tune $K_c$ by the interaction strength to favor the SC phase in the $\Delta_m<0$ region for the emergence of MKPs. 

\par Originating from the pseudospin sector, the stability of the MKP is protected by the gap in this sector. Given the decoupling between the charge and pseudospin sectors, any perturbations in the charge sector will not destabilize the MKP. Furthermore, deep in the phase where $m$ is relevant, the (RG irrelevant) operators coupling the two sectors will exhibit exponentially decaying correlation functions due to the gapped pseudospin sector, and thus keeping the MKP localized \cite{PhysRevLett.113.256405}.
 
\par We now examine the conservation of fermion parity and the ground-state degeneracy of Eq.~\eqref{eq:23}. Although the presence of pairing violates the fermion number conservation, it maintains the $\mathbb{Z}_2$ fermion parity,
\begin{equation}\label{eq:41}
    (-1)^{N_F} 
    = (-1)^{\int dr \big[ \mathcal{R}_{\tau}^\dagger(r)\mathcal{R}_{\tau}(r) + \mathcal{L}_{\tau}^\dagger(r)\mathcal{L}_{\tau}(r) \big] } 
    = e^{i \int dr \partial_r\varphi_\tau'(r)},
\end{equation}
which is related to the fermion number difference between the two time-reversal sectors. In the strong-coupling limit where $\vartheta_\tau'$ is pinned at its classical values, one can construct the ground states, $\ket{G_\pm}$, respecting the fermion parity in the bosonic language \cite{SM},
\begin{equation}
\label{eq:29}
\ket{G_\pm} \equiv
\begin{cases}
    \frac{1}{\sqrt{2}}(\ket{\frac{\pi}{2}} \pm \ket{\frac{3\pi}{2}}), & \text{ for } \Delta_m>0    ,
    \\ 
    \frac{1}{\sqrt{2}}(\ket{0} \pm \ket{\pi}), & \text{ for } \Delta_m<0.
\end{cases}
\end{equation}
Here, $\ket{\vartheta_0}$ is the eigenstate of $\vartheta_\tau'$ such that $\vartheta_\tau'\ket{\vartheta_0}=\vartheta_0\ket{\vartheta_0}$ and $(-1)^{N_F}\ket{G_\pm} = \pm\ket{G_\pm}$. Since the ground-state degeneracy is protected by the fermion parity conservation, the degenerate MZMs in a MKP will not be mixed by local perturbations and can be used to form qubits \cite{PhysRevB.84.195436, PhysRevLett.111.056402}.


\begin{figure}[t]
  \centering
  \centering
    \includegraphics[width=\linewidth]{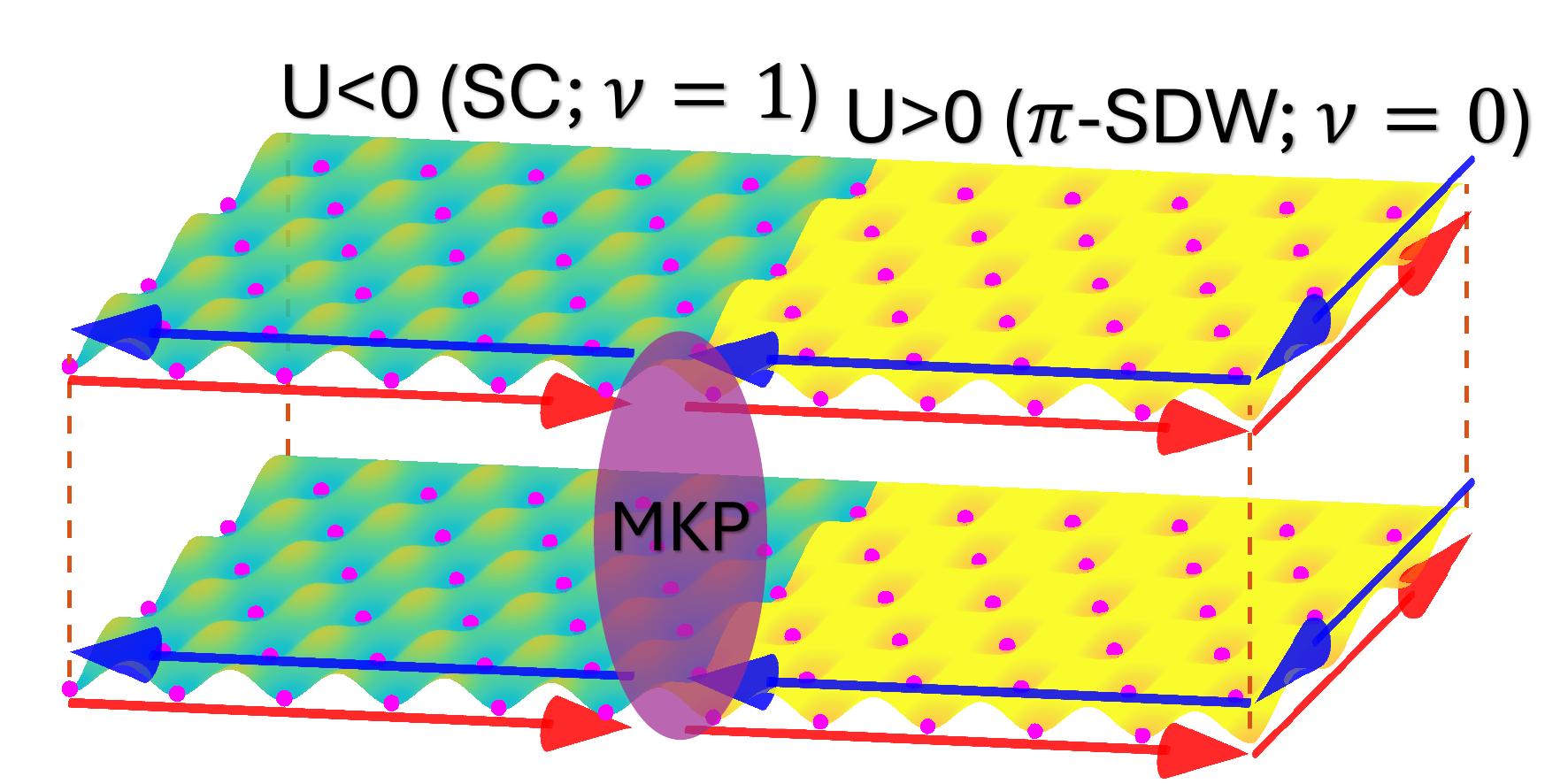}
  \caption{Synthetic double helical edge states of SCI and MKP in an optical lattice. A synthetic single SCI edge with $C_s=2$ can be achieved by tunnel-coupling the edges of two SCI layers with $C_s=1$. With intra-edge Hubbard interaction $U$ controlled through the Feshbach resonance,  a (green) region with $U<0$ and the other (yellow) with $U>0$ are created. MKP appears at the interface separating the SC and $\pi$-SDW phases with distinct $\mathbb{Z}_2$ invariants labeled by $\nu$ given in Eq.~\eqref{eq:26}. 
  }
  \label{fig:04}
\end{figure}

\paragraph{Experimental Realization---}While the general model in Eqs.~\eqref{eq:02}--\eqref{eq:03} is applicable broadly, we now discuss specific realizations. Motivated by their high controllability, we explore the cold-atom systems. To synthesize the double helical edge states, we consider the setup in FIG.~\ref{fig:04}, where two layers of SCI with $|C_s|=1$ can be stabilized through laser-induced synthetic gauge fields, as proposed theoretically~\cite{PhysRevLett.109.205303, PhysRevLett.105.255302, PhysRevLett.111.225301, Scheurer2015, PhysRevLett.97.240401, PhysRevLett.109.145301} and demonstrated experimentally in Ref.~\cite{PhysRevLett.111.185301}. Upon tunnel coupling, at these SCI edges~\cite{PhysRevLett.125.010403, Gall2021, samland2024thermodynamicsdensityfluctuationsbilayer, Meng2023}, one can effectively realize a single SCI edge with $|C_s|=2$. This synthetic configuration resembles the low-energy states observed in solid-state systems, including bilayers $\text{Bi}_\text{4}\text{Br}_\text{4}$~\cite{PhysRevB.109.155143} and coupled HgTe quantum wells \cite{PhysRevB.85.125309, PhysRevB.101.241302, PhysRevB.103.L201115, PhysRevB.93.235436, Ferreira2022, Krishtopenko2016}. Alternatively, double helical edge states could be directly established in cold-atom systems by simulating the tight-binding models of SCI monolayers with $|C_s|=2$ \cite{shulman2010robust, PhysRevB.108.245103, PhysRevB.94.235111, PhysRevB.110.035125, PhysRevB.104.L201110, PhysRevB.110.035161}.

Once the double helical edge state is stabilized, one can create a spatially varying Hubbard interaction $U$ within the edges using Feshbach resonances~\cite{RevModPhys.82.1225, PhysRevLett.93.050401, PhysRevA.92.053612, Mitra2018, PhysRevA.73.042705, PhysRevLett.91.020402, PhysRevA.85.051602}, with the sign of $U$ changing across different regions. In this setup, we have $\text{sign}(U)=\text{sign}(m)=\text{sign}(1-K_c)$~\cite{SM}, leading to SC and $\pi$-SDW phases depending on the sign of $U$. A MKP thus emerges at the interface between the $U<0$ and $U>0$ regions. Importantly, this scheme does not require a particularly strong interaction strength. Additionally, owing to their topological nature, these MKPs do not require a sharply defined interface, and their wave functions remain localized even with a smoother interface~\cite{SM}. These features will simplify the creation of interfaces in cold atom systems through the zero crossing of the Feshbach resonance curve with a moderate magnetic field gradient \cite{PhysRevA.73.042705, PhysRevLett.91.020402, PhysRevA.85.051602}.

\begin{figure}[t]
  \centering
  \centering
    \includegraphics[width=\linewidth]{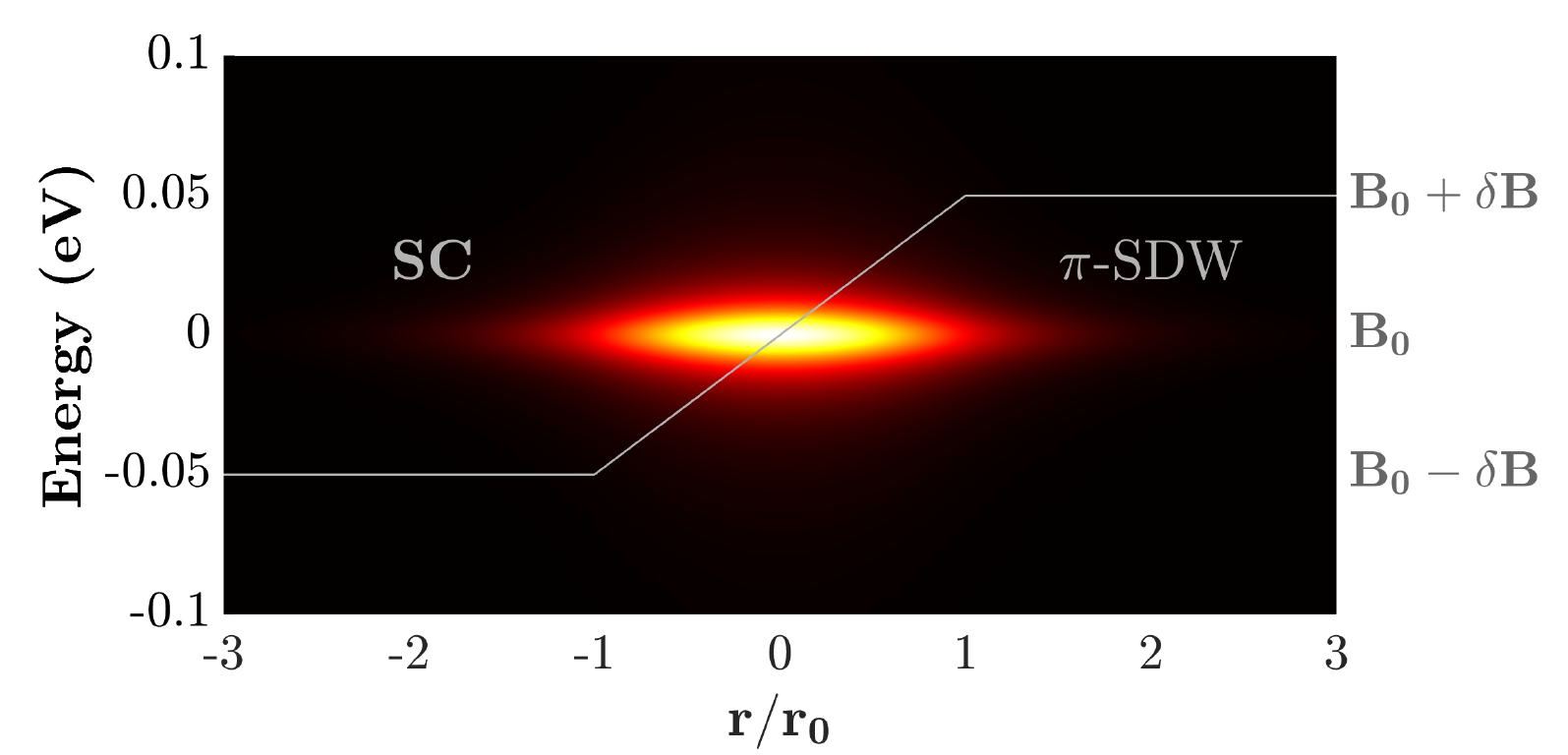}
  \caption{Computed LDOS of the MKP located around the interface ($r=0$) in the setup of FIG.~\ref{fig:04}. The overlaid gray line represents the spatial dependence of the applied magnetic field used for adjusting the interaction strength. Here, $B_0$ denotes the field strength at the zero crossing on the Feshbach resonance curve, with $\delta B$ defining the nonuniform field. Without loss of generality, we set $\hbar v = |\Delta_m|r_0$ and the spectral line broadening $\Gamma=10^{-2}$ eV~\cite{SM}.
  }
  \label{fig:05}
\end{figure}

Since MKPs manifest themselves as zero-energy modes, one could detect the local density of states (LDOS) through spatially-resolved radio frequency spectroscopy \cite{PhysRevLett.99.090403, PhysRevA.83.061604, PhysRevA.85.021603, PhysRevA.86.063604, PhysRevA.91.023610}. Following Ref.~\cite{PhysRevA.85.021603}, we compute the LDOS near an interface at $r=0$; see SM for details \cite{SM}. Figure.~\ref{fig:05} shows that the resulting LDOS has a Gaussian profile with the localization length, 
\begin{equation}
\label{eq:13}
    \xi_\text{loc} = \sqrt{\frac{\hbar v}{|\Delta_m|} r_0},
\end{equation}
where the length $r_0$ quantifies the interface's smoothness. This observable would provide a spectroscopic signature of the MKP.

\par We emphasize that our scheme relies solely on the edge theory, so that the locations of the MKP can be controlled through a magnetic field gradient. This simplifies detection compared to the schemes that use line solitons \cite{PhysRevLett.123.060402}. Our scheme also
eliminates the need for induced spin-orbit coupling as required in some proposals~\cite{PhysRevA.103.013307, Ye2017}. Although we have focused on the edges of SCIs and an experimentally accessible cold-atom platform~\cite{PhysRevLett.111.185301}, we expect our scheme to be applicable also to the edges of the MCIs. 

\paragraph{Acknowledgement---}We are grateful to Professor Ivana Dimitrova for important discussions in connection with cold-atom physics. The work at Northeastern University was supported by the National Science Foundation through the Expand-QISE award NSF-OMA-2329067 and benefited from the resources of Northeastern University’s Advanced Scientific Computation Center, the Discovery Cluster, and the Massachusetts Technology Collaborative award MTC-22032. C.-H.H. acknowledges the support from the National Science and Technology Council (NSTC), Taiwan through Grant No.~NSTC-112-2112-M-001-025-MY3.

\bibliography{ref}

\setcounter{equation}{0}
\setcounter{figure}{0}
\setcounter{table}{0}

\renewcommand{\theequation}{S\arabic{equation}}
\renewcommand{\thefigure}{S\arabic{figure}}
\renewcommand{\thetable}{S\arabic{table}}
\renewcommand{\bibnumfmt}[1]{[S#1]}
\renewcommand{\citenumfont}[1]{S#1}
\newcommand{\bk}{\boldsymbol\kappa}

\newcommand{\beginsupplement}{%
  \setcounter{equation}{0}
  \renewcommand{\theequation}{S\arabic{equation}}%
  \setcounter{table}{0}
  \renewcommand{\thetable}{S\arabic{table}}%
  \setcounter{figure}{0}
  \renewcommand{\thefigure}{S\arabic{figure}}%
  \setcounter{section}{0}
  \renewcommand{\thesection}{S\Roman{section}}%
  \setcounter{subsection}{0}
  \renewcommand{\thesubsection}{S\Roman{section}.\Alph{subsection}}%
}

\clearpage
\pagebreak
\widetext
\begin{center}
\textbf{\large Supplemental Materials: Engineering Majorana Kramers Pairs In Synthetic High Spin Chern Insulators}
\end{center}
\tableofcontents

\section{S1. Effects of interactions and bosonization}
\par Bosonic representations of the right and left moving modes in Eq.~(1) of the main text are:
\begin{equation}\label{eq:S1}
\begin{split}
    R_\mu(r) & = \frac{\kappa_R}{\sqrt{2\pi a}}e^{i[\vartheta_{\mu}(r)-\varphi_{\mu}(r)]}, \\
    L_\mu(r) & = \frac{\kappa_L}{\sqrt{2\pi a}}e^{i[\vartheta_{\mu}(r)+\varphi_{\mu}(r)]},
\end{split}
\end{equation}
where \(a\to 0\) is the length cutoff of the low-energy theory, \(\mu=1,2\) denotes the time-reversal sector, \(\kappa_{R(L)}\) is the Klein factor \cite{PhysRevLett.121.196801}, and the bosonic fields satisfy the commutation relation $[\varphi_i(r),\vartheta_j(r')]=-i\frac{\pi}{2}\text{sgn}(r-r')\delta_{ij}$ for $i,j=1,2$ \cite{book:Giamarchi}. Since all the right (left)-moving modes carry up (down) spin, we suppress the spin index in Eq.~\eqref{eq:S1} for simplicity. Using standard technique \cite{book:Giamarchi}, we then bosonize the Hamiltonian in Eq.~(2) of the main text to obtain: 
\begin{align}
    H_0 = & \int dr\,\frac{\hbar v}{2\pi} \sum_{\mu=1,2} (\partial_r\varphi_\mu(r))^2 + (\partial_r\vartheta_\mu(r))^2 .
\end{align}

\par In the following, we consider general short-range density-density and current-current interactions, described by
\begin{equation}\label{eq:S2}
    H_\text{int} = \frac{1}{2}\int drdr' V(r-r')\rho(r)\rho(r') . 
\end{equation}
Here, $\rho(r)$ denotes charge density, and $V(r-r')$ is the potential related to the short-range density-density interactions:
\begin{align*}
    \rho(r) = & \sum_{\mu=1,2} R^\dagger_\mu(r) R_\mu(r) + L^\dagger_\mu(r) L_\mu(r)+ [ R^\dagger_1(r) R_2(r) e^{-i2k_0r} +  L^\dagger_1(r) L_2(r) e^{i2k_0r} + \text{H.c.} ],
\end{align*}
where $k_0\equiv(k_1+k_2)/2$. The density-density interaction can thus be separated into two terms, $H_\text{int}=H_\text{int}^{(1)}+ H_\text{int}^{(2)}$, where $H_\text{int}^{(1)}$ involves the zero-momentum Fourier component $V_0$ of the interaction:
\begin{equation}\label{eq:S3}
\begin{split}
    H_\text{int}^{(1)} & = \frac{1}{2}\int drdr' \sum_{\mu,\nu=1,2}V(r-r')[R^\dagger_\mu(r) R_\mu(r) + L^\dagger_\mu(r) L_\mu(r)][R^\dagger_\nu(r') R_\nu(r') + L^\dagger_\nu(r') L_\nu(r')]
    \\ & = \frac{1}{2}\int dRdl \sum_{\mu,\nu=1,2}V(l)[R^\dagger_\mu(R+\frac{l}{2}) R_\mu(R+\frac{l}{2}) + L^\dagger_\mu(R+\frac{l}{2}) L_\mu(R+\frac{l}{2})][R^\dagger_\nu(R-\frac{l}{2}) R_\nu(R-\frac{l}{2}) + L^\dagger_\nu(R-\frac{l}{2}) L_\nu(R-\frac{l}{2})]
    \\ & \cong \frac{V_0}{2}\int dR\sum_{\mu,\nu=1,2}[:R^\dagger_\mu(R) R_\mu(R): + :L^\dagger_\mu(R) L_\mu(R):][:R^\dagger_\nu(R) R_\nu(R): + :L^\dagger_\nu(R) L_\nu(R):],
\end{split}
\end{equation}
and $H_\text{int}^{(2)}$ involves the finite-momentum component $V_{2k_0}$: 
\begin{equation}\label{eq:S4}
\begin{split}
    H_\text{int}^{(2)} & \cong \frac{1}{2}\int drdr' V(r-r')\{ e^{-i2k_0(r-r')}[R^\dagger_1(r) R_2(r) + L^\dagger_2(r) L_1(r)][R^\dagger_2(r') R_1(r') + L^\dagger_1(r') L_2(r')] + \text{H.c.} \}
    \\ & = \frac{1}{2}\int drdr' V(r-r')\{ e^{-i2k_0(r-r')}[R^\dagger_1(r) R_2(r)R^\dagger_2(r') R_1(r') + L^\dagger_2(r) L_1(r)L^\dagger_1(r') L_2(r')] + \text{H.c.} \}
    \\ & \quad + \frac{1}{2}\int drdr' V(r-r')\{ e^{-i2k_0(r-r')}[R^\dagger_1(r) R_2(r)L^\dagger_1(r') L_2(r') + L^\dagger_2(r) L_1(r)R^\dagger_2(r') R_1(r')] + \text{H.c.} \}
    \\ & = \frac{1}{2}\int drdr' V(r-r')\{ e^{-i2k_0(r-r')}[:R^\dagger_1(r)R_1(r'): + \langle R^\dagger_1(r)R_1(r')\rangle_0][-:R^\dagger_2(r)R_2(r'): - \langle R^\dagger_2(r)R_2(r')\rangle_0] + \text{H.c.} \}
    \\ & \quad + \frac{1}{2}\int drdr' V(r-r')\{ e^{-i2k_0(r-r')}[:L^\dagger_1(r)L_1(r'): + \langle L^\dagger_1(r)L_1(r')\rangle_0][-:L^\dagger_2(r)L_2(r'): - \langle L^\dagger_2(r)L_2(r')\rangle_0] + \text{H.c.} \}
    \\ & \quad + \frac{1}{2}\int drdr' V(r-r')\{ e^{-i2k_0(r-r')}[:R^\dagger_1(r) R_2(r)::L^\dagger_1(r') L_2(r'): + :L^\dagger_2(r) L_1(r)::R^\dagger_2(r') R_1(r'):] + \text{H.c.} \}
    \\ & = \frac{1}{2}\int dRdl V(l)\{ e^{-i2k_0(l)}[:R^\dagger_1(R+\frac{l}{2})R_1(R-\frac{l}{2}): + \frac{1}{\pi l}][-:R^\dagger_2(R+\frac{l}{2})R_2(R-\frac{l}{2}): - \frac{1}{\pi l}] + \text{H.c.} \}
    \\ & \quad + \frac{1}{2}\int dRdl V(l)\{ e^{-i2k_0(l)}[:L^\dagger_1(R+\frac{l}{2})L_1(R-\frac{l}{2}): + \frac{1}{\pi l}][-:L^\dagger_2(R+\frac{l}{2})L_2(R-\frac{l}{2}): - \frac{1}{\pi l}] + \text{H.c.} \}
    \\ & \quad + \frac{1}{2}\int dRdl V(l)\{ e^{-i2k_0(l)}[:R^\dagger_1(R+\frac{l}{2}) R_2(R+\frac{l}{2})::L^\dagger_1(R-\frac{l}{2}) L_2(R-\frac{l}{2}): 
    \\ & \quad \quad \quad \quad \quad \quad \quad \quad \quad \quad \quad \quad \quad + :L^\dagger_2(R+\frac{l}{2}) L_1(R+\frac{l}{2})::R^\dagger_2(R-\frac{l}{2}) R_1(R-\frac{l}{2}):] + \text{H.c.} \}
    \\ & \cong \frac{1}{2}\int dRdl V(l)\{ e^{-i2k_0(l)}[-:R^\dagger_1(R)R_1(R)::R^\dagger_2(R)R_2(R):] + \text{H.c.} \} 
    \\ & \quad - \frac{1}{2}\int dRdl V(l)\{\frac{e^{-i2k_0(l)}}{2\pi}\sum_{\mu=1,2}[(\partial_RR^\dagger_\mu(R))R_\mu(R) - R^\dagger_\mu(R)(\partial_RR_\mu(R))] + \text{H.c.} \}
    \\ & \quad + \frac{1}{2}\int dRdl V(l)\{ e^{-i2k_0(l)}[-:L^\dagger_1(R)L_1(R)::L^\dagger_2(R)L_2(R):] + \text{H.c.} \} 
    \\ &  \quad - \frac{1}{2}\int dRdl V(l)\{\frac{e^{-i2k_0(l)}}{2\pi}\sum_{\mu=1,2}[(\partial_RL^\dagger_\mu(R))L_\mu(R) - L^\dagger_\mu(R)(\partial_RL_\mu(R))] + \text{H.c.} \}
    \\ & \quad + \frac{1}{2}\int dRdl V(l)\{ e^{-i2k_0(l)}[:R^\dagger_1(R) R_2(R)::L^\dagger_1(R) L_2(R): + :L^\dagger_2(R) L_1(R)::R^\dagger_2(R) R_1(R):] + \text{H.c.} \}
    \\ & = V_{2k_0}\int dR [-:R^\dagger_1(R)R_1(R)::R^\dagger_2(R)R_2(R):-:L^\dagger_1(R)L_1(R)::L^\dagger_2(R)L_2(R):]
    \\ & \quad + V_{2k_0}\int dR [:R^\dagger_1(R) R_2(R)::L^\dagger_1(R) L_2(R): + \text{H.c.} ],
\end{split}
\end{equation}
Here, we neglect the fast-oscillating terms ($\propto e^{i2k_0(r+r')}$), the constant terms, the terms renormalizing the chemical potential, and all the irrelevant terms from dimensional analysis. We also apply Wick's theorem, and keep terms only up to $\mathcal{O}(l)$ throughout the derivation in view of the assumption of slow-varying fields \cite{PhysRevB.61.13410, PhysRevB.102.165147}, where we assume $R\equiv(r+r')/2$ and $l\equiv r-r'$, and $R\gg l$. Then, $H_\text{int}^{(1)}$ and $ H_\text{int}^{(2)}$ can be further bosonized into:
\begin{equation}\label{eq:S5}
\begin{split}
    H_\text{int}^{(1)} & = \frac{V_0}{2\pi^2}\int dr\sum_{\mu,\nu=1,2}[\partial_r\varphi_\mu(r)][\partial_r\varphi_\nu(r)] 
    \\ & = \frac{V_0}{4\pi^2}\int dr\sum_{\xi,\zeta=\pm1}[\partial_r(\varphi_c(r)+\xi\varphi_\tau(r))][\partial_r(\varphi_c(r)+\zeta\varphi_\tau(r))]
    \\ & = \frac{V_0}{\pi^2}\int dr [\partial_r\varphi_c(r)]^2,
\end{split}
\end{equation}
and
\begin{equation}\label{eq:S5.5}
\begin{split}
    H_\text{int}^{(2)} & = \frac{-V_{2k_0}}{4\pi^2}\int dr \partial_r[\vartheta_1(r)-\varphi_1(r)]\partial_r[\vartheta_2(r)-\varphi_2(r)] + \partial_r[\vartheta_1(r)+\varphi_1(r)]\partial_r[\vartheta_2(r)+\varphi_2(r)]
    \\ & \quad + \frac{V_{2k_0}}{4\pi^2a^2}\int dre^{i[\varphi_1(r)-\vartheta_1(r)+\vartheta_2(r)-\varphi_2(r)]}e^{i[-\varphi_1(r)-\vartheta_1(r)+\vartheta_2(r)+\varphi_2(r)]} + \text{H.c.}
    \\ & = \frac{-V_{2k_0}}{2\pi^2}\int dr \partial_r[\vartheta_1(r)]\partial_r[\vartheta_2(r)] + \partial_r[\varphi_1(r)]\partial_r[\varphi_2(r)]
    \\ & \quad + \frac{V_{2k_0}}{2\pi^2a^2}\int dr\cos[2\vartheta_1(r)-2\vartheta_2(r)]
    \\ & = \frac{-V_{2k_0}}{4\pi^2}\int dr \partial_r[\vartheta_c(r)+\vartheta_\tau(r)]\partial_r[\vartheta_c(r)-\vartheta_\tau(r)] + \partial_r[\varphi_c(r)+\varphi_\tau(r)]\partial_r[\varphi_c(r)-\varphi_\tau(r)]
    \\ & \quad + \frac{V_{2k_0}}{2\pi^2a^2}\int dr\cos[2\sqrt{2}\vartheta_\tau(r)]
    \\ & = \frac{V_{2k_0}}{4\pi^2}\int dr -[\partial_r\varphi_c(r)]^2 + [\partial_r\varphi_\tau(r)]^2 - [\partial_r\vartheta_c(r)]^2 + [\partial_r\vartheta_\tau(r)]^2
    \\ & \quad + \frac{V_{2k_0}}{2\pi^2a^2}\int dr\cos[2\sqrt{2}\vartheta_\tau(r)].
\end{split}
\end{equation}
Here, we define the charge and pseudospin degree of freedom labeled by the indices $c$ and $\tau$, respectively:
\begin{equation}\label{eq:S6}
    \begin{pmatrix}\varphi_c(r) \\ \varphi_\tau(r) \end{pmatrix} \equiv \frac{1}{\sqrt{2}}\begin{pmatrix} 1 & 1 \\ 1 & -1 \end{pmatrix}\begin{pmatrix}\varphi_1(r) \\ \varphi_2(r) \end{pmatrix},\quad \begin{pmatrix}\vartheta_c(r) \\ \vartheta_\tau(r) \end{pmatrix} \equiv \frac{1}{\sqrt{2}}\begin{pmatrix} 1 & 1 \\ 1 & -1 \end{pmatrix}\begin{pmatrix}\vartheta_1(r) \\ \vartheta_2(r) \end{pmatrix},
\end{equation}
which satisfies the commutation relation $[\varphi_\mu(r),\vartheta_\nu(r')]=-i\frac{\pi}{2}\text{sgn}(r-r')\delta_{\mu\nu}$ for $\mu,\nu=c,\tau$. To prove this commutation relation, note that Eq.~\eqref{eq:S6} can be expressed as $\varphi_\mu(r) = (U)_{\mu i}\varphi_i(r)$ and $\vartheta_\mu(r) = (U)_{\mu i}\vartheta_i(r)$ for $\mu=c,\tau$ and $i=1,2$, where $U=\frac{1}{\sqrt{2}}\begin{pmatrix} 1 & 1 \\ 1 & -1 \end{pmatrix}$. Then, since $U$ is orthogonal, one obtains 
\begin{equation}
\begin{split}
[\varphi_\mu(r),\vartheta_\nu(r')] & = (U)_{\mu i}(U)_{\nu j}[\varphi_i(r),\vartheta_j(r')] 
\\ & = -i\frac{\pi}{2}\text{sgn}(r-r')(U)_{\mu i}(U)_{\nu j}\delta_{ij}
\\ & = -i\frac{\pi}{2}\text{sgn}(r-r')(U)_{\mu i}(U^T)_{i\nu}
\\ & = -i\frac{\pi}{2}\text{sgn}(r-r')(UU^T)_{\mu\nu}
\\ & = -i\frac{\pi}{2}\text{sgn}(r-r')\delta_{\mu\nu}.
\end{split}
\end{equation}
The charge and pseudospin sectors represent the symmetric and anti-symmetric sectors, respectively, as described in the literature~\cite{PhysRevLett.103.166403, PhysRevB.109.155119, PhysRevB.99.045125, PhysRevB.98.115146, PhysRevB.93.235436, PhysRevB.86.165121}. The nomenclature here reflects the nature of the collective excitations present within these sectors, a point to which we will return below.

\par The Hamiltonian for a general short-range current-current interaction can be expressed as:
\begin{equation}\label{eq:S7}
    \tilde{H}_\text{int} = \frac{1}{2}\int drdr' \tilde{V}(r-r')J(r)J(r') .
\end{equation}
Here, $J(r)$ is the current density and $\tilde{V}(r-r')$ is the potential for the short-range current-current interaction. Due to the helical nature of the edge states, the current density, $J(r)$, is equivalent to the spin density, $\rho_s(r)$, up to a proportionality constant, see equations above. Therefore, the current-current interaction can be regarded as the spin-density-spin-density interaction in helical systems. More specifically:
\begin{align*}
    J(r) = & \sum_{\mu=1,2} R^\dagger_\mu(r) R_\mu(r) - L^\dagger_\mu(r) L_\mu(r)+ [ R^\dagger_1(r) R_2(r) e^{-i2k_0r} -  L^\dagger_1(r) L_2(r) e^{i2k_0r} + \text{H.c.} ].
\end{align*}
The current-current interaction can be separated into two terms like the density-density interaction, $\tilde{H}_\text{int}=\tilde{H}_\text{int}^{(1)} + \tilde{H}_\text{int}^{(2)}$, where $\tilde{H}_\text{int}^{(1)}$ involves the zero-momentum Fourier component $\tilde{V}_0$ of the interaction:
\begin{equation}\label{eq:S9}
\begin{split}
    \tilde{H}_\text{int}^{(1)} & = \frac{1}{2}\int drdr' \sum_{\mu,\nu=1,2}\tilde{V}(r-r')[R^\dagger_\mu(r) R_\mu(r) - L^\dagger_\mu(r) L_\mu(r)][R^\dagger_\nu(r') R_\nu(r') - L^\dagger_\nu(r') L_\nu(r')]
    \\ & \cong \frac{\tilde{V}_0}{2}\int dR\sum_{\mu,\nu=1,2}[:R^\dagger_\mu(R) R_\mu(R): - :L^\dagger_\mu(R) L_\mu(R):][:R^\dagger_\nu(R) R_\nu(R): - :L^\dagger_\nu(R) L_\nu(R):],
\end{split}
\end{equation}
and $\tilde{H}_\text{int}^{(2)}$ involves the zero-momentum Fourier component $\tilde{V}_{2k_0}$ of the interaction:
\begin{equation}\label{eq:S10}
\begin{split}
    \tilde{H}_\text{int}^{(2)} & = \frac{1}{2}\int drdr' \tilde{V}(r-r')\{ e^{-i2k_0(r-r')}[R^\dagger_1(r) R_2(r) - L^\dagger_2(r) L_1(r)][R^\dagger_2(r') R_1(r') - L^\dagger_1(r') L_2(r')] + \text{H.c.} \}
    \\ & = \frac{1}{2}\int drdr' \tilde{V}(r-r')\{ e^{-i2k_0(r-r')}[R^\dagger_1(r) R_2(r)R^\dagger_2(r') R_1(r') + L^\dagger_2(r) L_1(r)L^\dagger_1(r') L_2(r')] + \text{H.c.} \}
    \\ & \quad - \frac{1}{2}\int drdr' \tilde{V}(r-r')\{ e^{-i2k_0(r-r')}[R^\dagger_1(r) R_2(r)L^\dagger_1(r') L_2(r') + L^\dagger_2(r) L_1(r)R^\dagger_2(r') R_1(r')] + \text{H.c.} \}
    \\ & \cong \tilde{V}_{2k_0}\int dR [-:R^\dagger_1(R)R_1(R)::R^\dagger_2(R)R_2(R):-:L^\dagger_1(R)L_1(R)::L^\dagger_2(R)L_2(R):] 
    \\ & \quad - \tilde{V}_{2k_0}\int dR [:R^\dagger_1(R) R_2(R)::L^\dagger_1(R) L_2(R): + \text{H.c.} ].
\end{split}
\end{equation}
Then, $\tilde{H}_\text{int}^{(1)}$ and $\tilde{H}_\text{int}^{(2)}$ can be further bosonized into:
\begin{align}
    \tilde{H}_\text{int}^{(1)} & = \frac{\tilde{V}_0}{\pi^2}\int dr [\partial_r\vartheta_c(r)]^2,
    \\ \tilde{H}_\text{int}^{(2)} & = \frac{\tilde{V}_{2k_0}}{4\pi^2}\int dr \{ -[\partial_r\varphi_c(r)]^2 + [\partial_r\varphi_\tau(r)]^2 - [\partial_r\vartheta_c(r)]^2 + [\partial_r\vartheta_\tau(r)]^2 \} - \frac{\tilde{V}_{2k_0}}{2\pi^2a^2}\int dr\cos[2\sqrt{2}\vartheta_\tau(r)].
\end{align}

\par By incorporating the density-density and current-current interactions into the bosonized Hamiltonian, $H_0$, we obtain the Hamiltonian of a two-channel Luttinger liquid, $H_{\text{LL}}$ and a sine-Gordon-like mass term, $H_{\text{m}}$:
\begin{equation}\label{eq:S11}
\begin{split}
    H_{\text{LL}} & = \int dr\,\frac{\hbar}{2\pi}\sum_{\mu=c,\tau} u_\mu[K_\mu^{-1}(\partial_r\varphi_\mu(r))^2 + K_\mu(\partial_r\vartheta_\mu(r))^2],
    \\ H_{\text{m}} & =  \int dr\,\frac{V_{2k_0}-\tilde{V}_{2k_0}}{2(\pi a)^2}\cos[2\sqrt{2}\vartheta_\tau(r)].
\end{split}
\end{equation} 
Here, the corresponding Luttinger liquid parameters are:
\begin{equation}\label{eq:S12}
    K_c = \sqrt{\frac{1+\frac{1}{\pi\hbar v}[2\tilde{V}_0 - \frac{1}{2}(V_{2k_0} + \tilde{V}_{2k_0})]}{1+\frac{1}{\pi\hbar v}[2V_0 - \frac{1}{2}(V_{2k_0} + \tilde{V}_{2k_0})]}}\text{ and }K_\tau = 1,
\end{equation}
and the renormalized velocities are:
\begin{equation}\label{eq:S12.5}
    u_c = \frac{v}{K_c}\{1+\frac{1}{\pi\hbar v}[2\tilde{V}_0 - \frac{1}{2}(V_{2k_0} + \tilde{V}_{2k_0})]\} \text{ and }u_\tau=\frac{v}{K_\tau}[1+\frac{1}{2\pi\hbar v}(V_{2k_0} + \tilde{V}_{2k_0})].
\end{equation}

\section{S2. Renormalization-group (RG) flow equations}
\par  According to Eq.~\eqref{eq:S11}, the perturbative RG flow equations to the leading order contribution can be derived through a standard procedure \cite{book:Giamarchi, PhysRevB.97.045415}:
\begin{align}
    \frac{dK_\tau}{dl} = &  2\tilde{m}^2,         \label{eq:S13}
    \\ \frac{d\tilde{m}}{dl} = & 2(1-K_\tau^{-1})\tilde{m},  \label{eq:S14}
\end{align}
where $\tilde{m}\equiv m/\hbar v$ and $m\equiv V_{2k_0}-\tilde{V}_{2k_0}$. The flow trajectories are described by:
\begin{equation}\label{eq:S15}
    \tilde{m}^2 = 2[K_\tau - \ln(K_\tau)] + \text{constant}. 
\end{equation}
Note that around the non-interacting point, $K_\tau=1$, the RG flow equations restore the celebrated Kosterlitz RG flow equations \cite{J_M_Kosterlitz_1974}, which can be understood better if we define $\delta K_\tau\equiv2(1-K_\tau)$ and $y\equiv2\tilde{m}$. Then, to order $\mathcal{O}(\delta K_\tau)$, equations~\eqref{eq:S13} and \eqref{eq:S14} become:
\begin{align}
    \frac{d\delta K_\tau}{dl} = &  -y^2,        
    \\ \frac{dy}{dl} = & -\delta K_\tau y.  
\end{align}
Since $K_{\tau}$ monotonically increases under the RG flow according to Eq.~\eqref{eq:S13}, $\tilde{m}$ is relevant only if:
\begin{align}\label{eq:SRG1}
[\tilde{m}(l=0)]^2 - 2\{K_\tau(l=0) - \ln[K_\tau(l=0)]\} > -2, \, \text{ or } \, K_\tau(l=0) \geq 1 \text{ with } \tilde{m}(l=0)\neq0,
\end{align}
based on the solutions shown in Eq.~\eqref{eq:S15}. Otherwise, $\tilde{m}$ is irrelevant. The critical line for $\tilde{m}$ to be relevant corresponds to:
\begin{equation}\label{eq:SRG2}
[\tilde{m}(l=0)]^2 - 2\{K_\tau(l=0) - \ln[K_\tau(l=0)]\} = -2 \text{ with } K_\tau(l=0) \leq 1.
\end{equation}
Note that Eqs.~\eqref{eq:SRG1} and \eqref{eq:SRG2} are equivalent to Eq.~(6) of the main text.

\par As FIG.~2 of the main text shows, the RG flow equations described byEqs.~\eqref{eq:S13} and \eqref{eq:S14} have drain-like (source-like) fixed points $(K_\tau,\tilde{m})=(K_\tau^*,0)$ for some $K_\tau^*<1\,(K_\tau^*>1)$ and a saddle-point-like fixed point $(K_\tau,\tilde{m})=(1,0)$. Furthermore, since $K_\tau(l=0)=1$ according to Eq.~\eqref{eq:S12}, the mass term remains relevant as long as $\tilde{m}(l=0)\neq0$ in view of Eq.~\eqref{eq:SRG1}.

Importantly, since the RG flows that lead to a relevant mass term with $K_\tau(l=0)=1$ always pass through $K_\tau=2$ without closing the energy gap, see FIG. 2 of the main text), the interacting Hamiltonian density of the pseudospin sector with $m\neq0$ can always be adiabatically connected to the non-interacting Hamiltonian density in Eq.~9 of the main text, providing a solid basis for the emergence of the MKP in the strongly interacting region in practice. 

\section{S3. Quasi-long-range orders}
\par Here, we derive the bosonized expression of the operators that may lead to quasi-long-range orders discussed in the main text. The operator characterizing the instability of the $x$ component of the spin-density wave (SDW), $\Delta_{2q}^{(x)}$, is:
\begin{equation}
\begin{split}
    \Delta_{2q}^{(x)} & = \sum_{\substack{\mu=1,2 \\ \nu\neq\mu}} L_\mu^\dagger(r) R_\nu(r) + \text{H.c.}
    \\ & \sim \frac{1}{2\pi a}[ e^{-i[\varphi_1(r)+\vartheta_1(r)+\varphi_2(r)-\vartheta_2(r)]} + e^{-i[\varphi_2(r)+\vartheta_2(r)+\varphi_1(r)-\vartheta_1(r)]} - \text{H.c.} ]
     \\ & = \frac{1}{2\pi a}[ e^{-i\sqrt{2}[\varphi_c(r)+\vartheta_\tau(r)]} + e^{-i\sqrt{2}[\varphi_c(r)-\vartheta_\tau(r)]} - \text{H.c.} ]
     \\ & = \frac{1}{\pi a}[ e^{-i\sqrt{2}\varphi_c(r)}\cos[\sqrt{2}\vartheta_\tau(r)] - \text{H.c.} ]
    \\ & = \frac{-2i}{\pi a}\sin[\sqrt{2}\varphi_c(r)]\cos[\sqrt{2}\vartheta_\tau(r)],
\end{split}
\end{equation}
where the subscript denotes the nesting vector $2q$; that is, the momentum difference between $L_\mu$ and $ R_\nu$ as shown in FIG. 1 of the main text.
Similarly, the operator characterizing the instability of the y component of the SDW, $\Delta_{2q}^{(y)}$, is:
\begin{equation}
\begin{split}
    \Delta_{2q}^{(y)} & = \sum_{\substack{\mu=1,2 \\ \nu\neq\mu}} iL_\mu^\dagger(r) R_\nu(r) + \text{H.c.}
    \\ & \sim \frac{1}{2\pi a}\{ ie^{-i[\varphi_1(r)+\vartheta_1(r)+\varphi_2(r)-\vartheta_2(r)]} + ie^{-i[\varphi_2(r)+\vartheta_2(r)+\varphi_1(r)-\vartheta_1(r)]} - \text{H.c.} \}
    \\ &  = \frac{1}{2\pi a}\{ ie^{-i\sqrt{2}[\varphi_c(r)+\vartheta_\tau(r)]} + ie^{-i\sqrt{2}[\varphi_c(r)-\vartheta_\tau(r)]} - \text{H.c.} \}
    \\ &  = \frac{1}{\pi a}\{ ie^{-i\sqrt{2}\varphi_c(r)}\cos[\sqrt{2}\vartheta_\tau(r)] - \text{H.c.} \}
    \\ & = \frac{2i}{\pi a}\cos[\sqrt{2}\varphi_c(r)]\cos[\sqrt{2}\vartheta_\tau(r)].
\end{split}
\end{equation}
In addition to the in-plane components of the SDW, the out-of-plane component of the SDW has a nesting vector $2k_0$. The corresponding operator, $\Delta_{2k_0}^{(\text{z})}$, can be expressed as:
\begin{equation}
\begin{split}
    \Delta_{2k_0}^{(\text{z})} & =  R^\dagger_1(r) R_2(r) - L^\dagger_1(r) L_2(r) + \text{H.c.}
    \\ & \sim \frac{1}{2\pi a}\{ e^{i[-\varphi_1(r)+\vartheta_1(r)+\varphi_2(r)-\vartheta_2(r)]} 
     - e^{i[\varphi_1(r)+\vartheta_1(r)-\varphi_2(r)-\vartheta_2(r)]}\} + \text{H.c.}
    \\ & = \frac{1}{\pi a}\{ \cos[-\sqrt{2}\varphi_\tau(r) + \sqrt{2}\vartheta_\tau(r)] 
     - \cos[\sqrt{2}\varphi_\tau(r) + \sqrt{2}\vartheta_\tau(r)]  \}
    \\ & = \frac{2}{\pi a}\sin[\sqrt{2}\varphi_\tau(r)]\sin[\sqrt{2}\vartheta_\tau(r)].
\end{split}
\end{equation}

Besides the regular SDWs, we define the $\pi$-SDW orders, which are SDWs with a $\pi$-phase difference for nesting between different pairs of fermi points, resulting in a sign change upon interchanging the time-reversal sectors. The operator characterizing the instability of its $x$-component, $\Delta_{2q,\pi}^{(x)}$, is:
\begin{equation}
\begin{split}
    \Delta_{2q,\pi}^{(x)} & = L_1^\dagger(r) R_2(r) + \text{H.c.} - (1\leftrightarrow2)
    \\ & \sim \frac{1}{2\pi a}[ e^{-i[\varphi_1(r)+\vartheta_1(r)+\varphi_2(r)-\vartheta_2(r)]} - e^{-i[\varphi_2(r)+\vartheta_2(r)+\varphi_1(r)-\vartheta_1(r)]} - \text{H.c.} ]
     \\ & = \frac{1}{2\pi a}[ e^{-i\sqrt{2}[\varphi_c(r)+\vartheta_\tau(r)]} - e^{-i\sqrt{2}[\varphi_c(r)-\vartheta_\tau(r)]} - \text{H.c.} ]
     \\ & = \frac{1}{\pi a}[ -ie^{-i\sqrt{2}\varphi_c(r)}\sin[\sqrt{2}\vartheta_\tau(r)] - \text{H.c.} ]
    \\ & = \frac{-2i}{\pi a}\cos[\sqrt{2}\varphi_c(r)]\sin[\sqrt{2}\vartheta_\tau(r)].
\end{split}
\end{equation}
Similarly, the operator characterizing the instability of the $y$-component of $\pi$-SDW, $\Delta_{2q,\pi}^{(y)}$, is:
\begin{equation}
\begin{split}
    \Delta_{2q,\pi}^{(y)} & = iL_1^\dagger(r) R_2(r) + \text{H.c.} - (1\leftrightarrow2)
    \\ & \sim \frac{1}{2\pi a}\{ ie^{-i[\varphi_1(r)+\vartheta_1(r)+\varphi_2(r)-\vartheta_2(r)]} - ie^{-i[\varphi_2(r)+\vartheta_2(r)+\varphi_1(r)-\vartheta_1(r)]} - \text{H.c.} \}
    \\ &  = \frac{1}{2\pi a}\{ ie^{-i\sqrt{2}[\varphi_c(r)+\vartheta_\tau(r)]} - ie^{-i\sqrt{2}[\varphi_c(r)-\vartheta_\tau(r)]} - \text{H.c.} \}
    \\ &  = \frac{1}{\pi a}\{ e^{-i\sqrt{2}\varphi_c(r)}\sin[\sqrt{2}\vartheta_\tau(r)] - \text{H.c.} \}
    \\ & = \frac{-2i}{\pi a}\sin[\sqrt{2}\varphi_c(r)]\sin[\sqrt{2}\vartheta_\tau(r)].
\end{split}
\end{equation}
Given that $R_\mu\to L_\mu$, $L_\mu\to -R_\mu$, and $i\to-i$ through TRS, the $\pi$-SDWs undergo the transformation $\Delta_{2q,\pi}^{(x/y)}\to\Delta_{2q,\pi}^{(x/y)}$ under TRS, thus maintaining TRS. As an alternative, TRS of $\pi$-SDWs can be verified using bosonic fields, where under TRS, $\varphi_\mu\to\varphi_\mu$ and $\vartheta_\mu\to-\vartheta_\mu+\pi$, according to Eq.~\eqref{eq:S1}. This results in transformations: $\varphi_{c(\tau)}\to\varphi_{c(\tau)}$, $\vartheta_c\to-\vartheta_c+\sqrt{2}\pi$, and $\vartheta_\tau\to-\vartheta_\tau$, according to Eq.~\eqref{eq:S6}. Consequently, together with $i\to-i$ through TRS, the $\pi$-SDWs preserve TRS according to their bosonized expressions.

\par
\begin{figure}[ht]
  \centering
  \centering
    \includegraphics[width=0.75\linewidth]{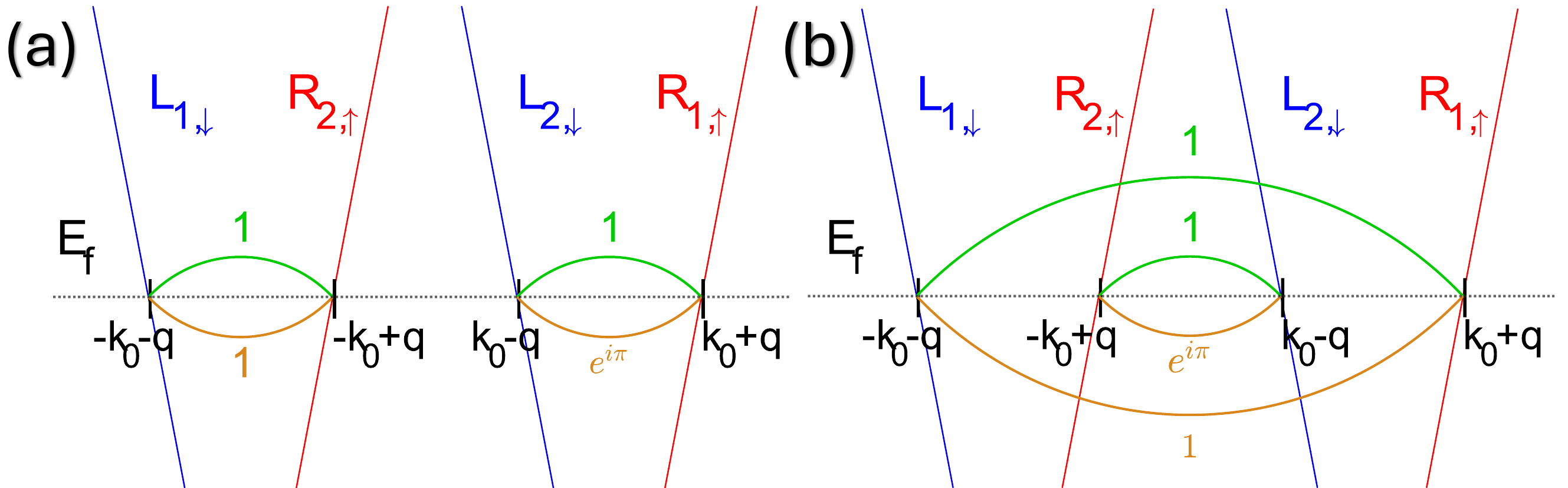}
  \caption{ Phase differences for (a) nesting between two pairs of fermi points in the SDW (green) and the $\pi$-SDW (orange), and (b) for paring between different time-reversal sectors in the SC (green) and the $\pi$-SC (orange) phases.  }
  \label{fig:S1}
\end{figure}

 The  operator characterizing the instability of the superconductivity (SC), $\Delta_{\text{SC}}$, is:
\begin{equation}
\begin{split}
    \Delta_{\text{SC}} & = \sum_{\mu=1,2} R_\mu^\dagger(r) L_\mu^\dagger(r) - L_\mu^\dagger(r) R_\mu^\dagger(r)
     \\ & \sim \frac{1}{\pi a}\{e^{i[\varphi_1(r)-\vartheta_1(r)-\vartheta_1(r)-\varphi_1(r)]}+(1\leftrightarrow2)\}
     \\ & = \frac{1}{\pi a}\{e^{-i2\vartheta_1(r)}+e^{-i2\vartheta_2(r)} \}
    \\ & = \frac{1}{\pi a}\{e^{-i\sqrt{2}[\vartheta_c(r)+\vartheta_\tau(r)]}+e^{-i\sqrt{2}[\vartheta_c(r)-\vartheta_\tau(r)]} \}
    \\ & = \frac{2}{\pi a}e^{-i\sqrt{2}\vartheta_c(r)}\cos[\sqrt{2}\vartheta_\tau(r)].
\end{split}
\end{equation}
We can also define the $\pi$-SC order, which is SC with out-of-phase pairing between different time-reversal sectors, analogous to the $\pi$-junction settings \cite{PhysRevLett.115.237001, Hsu_2021, PhysRevLett.100.096407, PhysRevLett.111.116402, PhysRevLett.122.126402, PhysRevLett.120.267002, PhysRevLett.129.227002, PhysRevB.97.014507, PhysRevB.89.220504}. The operator characterizing its instability, $\Delta_{\text{SC},\pi}$, is:
\begin{equation}
\begin{split}
    \Delta_{\text{SC},\pi} & = R_1^\dagger(r) L_1^\dagger(r) - L_1^\dagger(r) R_1^\dagger(r) - (1\leftrightarrow2)
     \\ & \sim \frac{1}{\pi a}\{e^{i[\varphi_1(r)-\vartheta_1(r)-\vartheta_1(r)-\varphi_1(r)]}-(1\leftrightarrow2)\}
     \\ & = \frac{1}{\pi a}\{e^{-i2\vartheta_1(r)}-e^{-i2\vartheta_2(r)} \}
    \\ & = \frac{1}{\pi a}\{e^{-i\sqrt{2}[\vartheta_c(r)+\vartheta_\tau(r)]}-e^{-i\sqrt{2}[\vartheta_c(r)-\vartheta_\tau(r)]} \}
    \\ & = \frac{-2i}{\pi a}e^{-i\sqrt{2}\vartheta_c(r)}\sin[\sqrt{2}\vartheta_\tau(r)].
\end{split}
\end{equation}
FIG.~\ref{fig:S1} schematically shows differences between SDW and $\pi$-SDW as well as  SC and $\pi$-SC.

We also consider several variations of the pair destiny wave (PDW) \cite{RevModPhys.87.457, annurev_PDW, PhysRevB.79.064515, PhysRevB.91.195102, PhysRevB.101.165133, PhysRevLett.113.256405}.   The operator characterizing the instability of the singlet pair density wave (PDW), $\Delta_{2q}^{(\text{sPDW})}$, is:
\begin{equation}
\begin{split}
\Delta_{2q}^{(\text{sPDW})} & = \sum_{\substack{\mu=1,2 \\ \nu\neq\mu}} R_\mu^\dagger(r) L_\nu^\dagger(r) - L_\nu^\dagger(r) R_\mu^\dagger(r)
\\ & \sim \frac{1}{\pi a}\{e^{i[\varphi_1(r)-\vartheta_1(r)-\vartheta_2(r)-\varphi_2(r)]}+ e^{i[\varphi_2(r)-\vartheta_2(r)-\vartheta_1(r)-\varphi_1(r)]}\}
\\ & = \frac{1}{\pi a}\{e^{i\sqrt{2}[\varphi_\tau(r)-\vartheta_c(r)]}+ e^{i\sqrt{2}[-\varphi_\tau(r)-\vartheta_c(r)]}\}
\\ & = \frac{2}{\pi a}e^{-i\sqrt{2}\vartheta_c(r)}\cos[\sqrt{2}\varphi_\tau(r)].
\end{split}
\end{equation}
Like the $\pi$-SDW, the operator characterizing the instability, $\Delta_{2q,\pi}^{(\text{sPDW})}$, for the $\pi$-singlet PDW is:
\begin{equation}
\begin{split}
\Delta_{2q,\pi}^{(\text{sPDW})} & = R_1^\dagger(r) L_2^\dagger(r) - L_2^\dagger(r) R_1^\dagger(r) - (1\leftrightarrow2)
\\ & \sim \frac{1}{\pi a}\{e^{i[\varphi_1(r)-\vartheta_1(r)-\vartheta_2(r)-\varphi_2(r)]} - e^{i[\varphi_2(r)-\vartheta_2(r)-\vartheta_1(r)-\varphi_1(r)]}\}
\\ & = \frac{1}{\pi a}\{e^{i\sqrt{2}[\varphi_\tau(r)-\vartheta_c(r)]} - e^{i\sqrt{2}[-\varphi_\tau(r)-\vartheta_c(r)]}\}
\\ & = \frac{2i}{\pi a}e^{-i\sqrt{2}\vartheta_c(r)}\sin[\sqrt{2}\varphi_\tau(r)].
\end{split}
\end{equation}
The operator characterizing the instability of the $x$-component of the triplet PDW, $\Delta_{2q}^{(\text{tPDW;x})}$, is:
\begin{equation}
\begin{split}
\Delta_{2q}^{(\text{tPDW;x})} & = \sum_{\substack{\mu=1,2 \\ \nu\neq\mu}} R_\mu^\dagger(r) R_\nu^\dagger(r) - L_\mu^\dagger(r) L_\nu^\dagger(r)
\\ & \sim \frac{1}{\pi a}\{e^{i[\varphi_1(r)-\vartheta_1(r)+\varphi_2(r)-\vartheta_2(r)]}- e^{-i[\varphi_1(r)+\vartheta_1(r)+\varphi_2(r)+\vartheta_2(r)]}\}
\\ & = \frac{1}{\pi a}\{e^{i\sqrt{2}[\varphi_c(r)-\vartheta_c(r)]}- e^{i\sqrt{2}[-\varphi_c(r)-\vartheta_c(r)]}\}
\\ & = \frac{2i}{\pi a}e^{-i\sqrt{2}\vartheta_c(r)}\sin[\sqrt{2}\varphi_c(r)],
\end{split}
\end{equation}
and the operator characterizing the instability of the $y$-component of the triplet PDW, $\Delta_{2q}^{(\text{tPDW;y})}$, is:
\begin{equation}
\begin{split}
\Delta_{2q}^{(\text{tPDW;y})} & = -i\sum_{\substack{\mu=1,2 \\ \nu\neq\mu}} R_\mu^\dagger(r) R_\nu^\dagger(r) + L_\mu^\dagger(r) L_\nu^\dagger(r)
\\ & \sim \frac{-i}{\pi a}\{e^{i[\varphi_1(r)-\vartheta_1(r)+\varphi_2(r)-\vartheta_2(r)]}+e^{-i[\varphi_1(r)+\vartheta_1(r)+\varphi_2(r)+\vartheta_2(r)]}\}
\\ & = \frac{-i}{\pi a}\{e^{i\sqrt{2}[\varphi_c(r)-\vartheta_c(r)]}+ e^{i\sqrt{2}[-\varphi_c(r)-\vartheta_c(r)]}\}
\\ & = \frac{-2i}{\pi a}e^{-i\sqrt{2}\vartheta_c(r)}\cos[\sqrt{2}\varphi_c(r)].
\end{split}
\end{equation}
And the operator characterizing the instability of the charge density wave (CDW), $\Delta_{2k_0}^{(\text{CDW})}$, is:
\begin{equation}
\begin{split}
\Delta_{2k_0}^{(\text{CDW})} & = R^\dagger_1(r) R_2(r) + L^\dagger_1(r) L_2(r) + \text{H.c.}
\\ & \sim \frac{1}{2\pi a}\{e^{i[\varphi_1(r)-\vartheta_1(r)+\vartheta_2(r)-\varphi_2(r)]} + e^{i[-\varphi_1(r)-\vartheta_1(r)+\vartheta_2(r)+\varphi_2(r)]} + \text{H.c.} \}
\\ & = \frac{1}{2\pi a}\{e^{i\sqrt{2}[\varphi_\tau(r)-\vartheta_\tau(r)]} + e^{i[-\varphi_\tau(r)-\vartheta_\tau(r)]} + \text{H.c.} \}
\\ & = \frac{1}{\pi a}\{e^{-i\sqrt{2}\vartheta_\tau(r)}\cos[\sqrt{2}\varphi_\tau(r)] + \text{H.c.} \}
\\ & = \frac{2}{\pi a}\cos[\sqrt{2}\vartheta_\tau(r)]\cos[\sqrt{2}\varphi_\tau(r)].
\end{split}
\end{equation}

\par As pointed out in the main text, the in-plane components of the SDWs and SCs develop quasi-long-range orders, which can host power-law-decaying correlation functions. When $\vartheta_\tau(r)$ is pinned due to $H_{\text{m}}$ in Eq.~\eqref{eq:S11}, thes correlations can be shown to scale as \cite{book:Giamarchi}:
\begin{align}
 \langle \Delta_{2q}^{(\pm)}(\mathfrak{r})\Delta_{2q}^{(\mp)}(0) \rangle & \sim |\mathfrak{r}|^{-K_c}
\\  \langle \Delta_{\text{SC}(,\pi)}(\mathfrak{r})\Delta_{\text{SC}(,\pi)}(0) \rangle & \sim |\mathfrak{r}|^{-K_c^{-1}}.
\end{align} 
Here, $\mathfrak{r}=(r,v\tau)$ with the imaginary time $\tau$ and $\Delta_{2q}^{(\pm)}\equiv\Delta_{2q(,\pi)}^{(x)}\pm i\Delta_{2q(,\pi)}^{(y)}$. The corresponding susceptibilities are:
\begin{align}
    \chi_{\text{SDW}} & \sim \omega^{K_c-2}         \label{eq:S16}
    \\ \chi_{\text{SC}} & \sim \omega^{K_c^{-1}-2},  \label{eq:S17}
\end{align}
suggesting that the quasi-long-range in-plane components of the SDW and SC orders are likely to form when $K_c<2$ and $K_c>\frac{1}{2}$, respectively, as a consequence of the divergent susceptibilities \cite{book:Giamarchi}. This will further gap out the charge sector. 

\begin{figure}[ht]
  \centering
  \centering
    \includegraphics[width=0.75\linewidth]{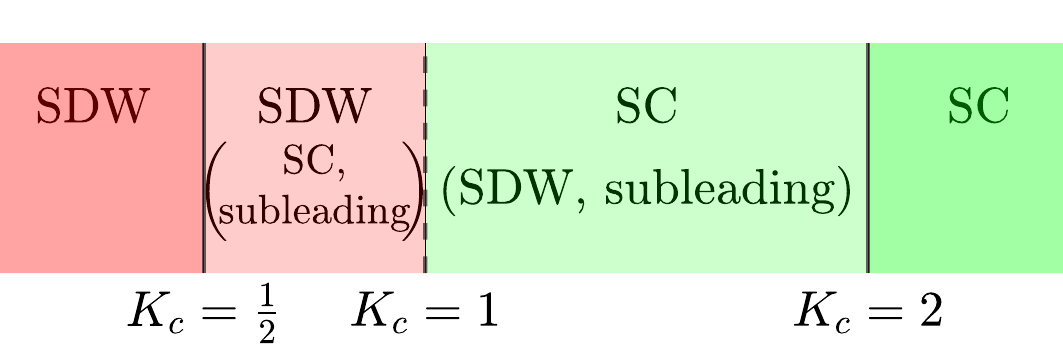}
  \caption{ Susceptibilities of the SDW and SC operators diverge for $K_c<2$ and $K_c>\frac{1}{2}$, respectively, leading to quasi-long-range orders. In the interval $\frac{1}{2}<K_c<2$, both SDW and SC orders coexist. However, for $K_c<1$ and $K_c>1$, the SDW and SC phases respectively overpower due to the more singular susceptibility.
  }
  \label{fig:S2}
\end{figure}

Within the region $\frac{1}{2}<K_c<2$, both the in-plane components of the SDWs and the SCs can develop a quasi-long-range ordering. However, with distinct scaling dimensions for $K_c \neq 1$, their susceptibilities in Eqs.~\eqref{eq:S16} and \eqref{eq:S17} have different singularities.  For $K_c<1$, the SDW susceptibility is more singular than in the SC case, whereas the situation is the opposite for $K_c>1$. A schematic diagram showing the corresponding phase diagram is shown in FIG.~\ref{fig:S2}.

Along the preceding lines, susceptibility of the correlation function of the triplet PDWs, $\Delta_{2q}^{(\text{tPDW;}\pm)}\equiv\Delta_{2q}^{(\text{tPDW;x})}\pm i\Delta_{2q}^{(\text{tPDW;y})}$ can be shown to be:
\begin{equation}
    \chi_{\text{tPDW;}\pm} \sim \omega^{K_c+K_c^{-1}-2},
\end{equation}
which does not display any order since $K_c+K_c^{-1}\geq2$.
Due to the fluctuation of $\varphi_\tau(r)$ when $\vartheta_\tau(r)$ is pinned, the out-of-plane component of the SDW, $\Delta_{2k_0}^{(\text{z})}$, the singlet PDWs, $\Delta_{2q}^{(\text{sPDW})}$ and $\Delta_{\pi;2q}^{(\text{sPDW})}$, and the CDW, $\Delta_{2k_0}^{(\text{CDW})}$, also do not stabilize any order.

\par Note that the in-plane components of the SDW are induced rather than the $\hat{z}$ component, despite the fact that the spin-$U(1)$ symmetry is not explicitly broken in the Hamiltonians in Eq.~\eqref{eq:S11}. This effect is better seen from an alternative expression of the Hamiltonian density of the mass term:
\begin{equation}\label{eq:09}
    H_\text{m}(r) \propto \Delta^{(+)}_{2q}(r)\Delta^{(-)}_{2q}(r) + \Delta^{(-)}_{2q}(r)\Delta^{(+)}_{2q}(r) ,
\end{equation}
up to a constant. This implies that the in-plane ($\hat{z}$) component of the SDW has nonzero (zero) correlation function when the mass term is relevant, which is consistent with the computation of the SDWs' characterizing operators.

\par We emphasize that in solid-state systems, where the repulsive Coulomb interaction dominates such that $\tilde{V}_0=\tilde{V}_{2k_0}=0$ and $V_0>V_{2k_0}>0$, we have $K_c<1$ for $m>0$ in view of Eqs.~\eqref{eq:S11} and \eqref{eq:S12}, which stabilizes  the $\pi$-SDW phase.

\section{S4. $\mathbb{Z}_2$ invariant of the emergent Bogoliubov-de Gennes Hamiltonian}
\par Here, we discuss the $\mathbb{Z}_2$ topological invariant for the Hamiltonian density in Eq.~(9) of the main text, which can be regularized into a topologically equivalent lattice model \cite{PhysRevB.82.115120}:
\begin{equation}\label{eq:S8}
    H_{\text{eff}}(k)=t\sin(k)\eta_0\sigma_3 - [\Delta_m + u(1-\cos(k))]\eta_2\sigma_2,
\end{equation}
where $|\Delta_m|<2u$. For an appropriate choice of $t$, the low-energy Hamiltonian density near $k=0$ of Eq.~\eqref{eq:S8} reduces to the Hamiltonian density in Eq.~(9) of the main text with the replacement $k\to -i\partial_r$. In view of the chiral symmetry, the Hamiltonian density of Eq.~\eqref{eq:S8} can thus be rewritten as \cite{PhysRevB.84.060504}:
\begin{equation}\label{eq:30}
     H_{\text{eff}}'(k) = \begin{pmatrix} 0 & f(k) \\ f^\dagger(k) & 0 \end{pmatrix} , 
\end{equation}
through a unitary transformation, $H_{\text{eff}}'(k)=UH_{\text{eff}}(k)U^\dagger$, where:
\begin{equation}
\begin{split}
    U & = \frac{1}{\sqrt{2}}\begin{pmatrix} \sigma_0 & \sigma_2 \\ \sigma_2 & -\sigma_0 \end{pmatrix},
    \\ \text{and } f(k) & = \begin{pmatrix} 0 & -it\sin(k)-[\Delta_m + u(1-\cos(k))] \\ -it\sin(k)+[\Delta_m + u(1-\cos(k))] & 0 \end{pmatrix}.
\end{split}
\end{equation}
Then, by diagonalizing Eq.~\eqref{eq:30}, we can construct the matrix:
\begin{equation}\label{eq:31}
    q(k) = \frac{1}{\lambda(k)}f(k) = \begin{pmatrix} 0 & -e^{i\theta(k)} \\ e^{-i\theta(k)} & 0 \end{pmatrix}.
\end{equation}
Here, $\lambda(k)$ is the absolute value of the eigenvalue of $f(k)$ and $e^{i\theta(k)}\equiv \frac{[\Delta_m + u(1-\cos(k))]+it\sin(k)}{\sqrt{[\Delta_m + u(1-\cos(k))]^2 + [t\sin(k)]^2}}$. The $\mathbb{Z}_2$ invariant, $\nu$, can be obtained from \cite{PhysRevB.84.060504}:
\begin{equation}
    (-1)^\nu = \frac{\text{Pf}[q^T(k=0)]}{\sqrt{\det[q^T(k=0)]}}\frac{\text{Pf}[q^T(k=\pi)]}{\sqrt{\det[q^T(k=\pi)]}}, 
\end{equation}
which straightforwardly yields the $\mathbb{Z}_2$ invariant of the Hamiltonian density in Eq.~\eqref{eq:S8}:
\begin{equation}\label{eq:33}
    \nu=\frac{1-\text{sign}(\Delta_m)}{2}.
\end{equation}

\section{S5. Zero-energy domain-wall states}
\par Here, we solve for the bound-state wave function at the domain wall between the $\Delta_m>0$ and $\Delta_m<0$ regions. To be concrete, we assume $\Delta_m=\Delta_1$ for $r>0$ and $\Delta_m=-\Delta_2$ for $r<0$, with $\Delta_1,\Delta_2>0$, and solve for the localized states with energy $E$ at $r=0$. By using the ansatz $e^{-\kappa_1 r}$ for $r>0$ and $e^{\kappa_2 r}$ for $r<0$, it can be shown from Eq.~(9) of the main text that:
\begin{equation}
    \hbar v\kappa_i = \sqrt{\Delta_i^2-E^2} \text{ for }i=1,2,
\end{equation}
and the corresponding bound state, $\Phi(r) = \theta(r)\Phi_+(r)e^{-\kappa_1 r} + \theta(-r)\Phi_-(r)e^{\kappa_2 r}$, is:
\begin{equation}\label{eq:35}
\begin{split}
    \Phi_+(r) & = \begin{pmatrix} \frac{\Delta_1}{E-i\hbar v\kappa_1} & 0 & 0 & 1 \end{pmatrix}^T
    \\ \Phi_-(r) & = \begin{pmatrix} \frac{-\Delta_2}{E+i\hbar v\kappa_2} & 0 & 0 & 1 \end{pmatrix}^T.
\end{split}
\end{equation}
Here, $\theta(r)$ is the Heaviside step function and the basis $\begin{pmatrix} \mathcal{R}_{\tau}(r) & \mathcal{L}_{\tau}(r) & \mathcal{R}_{\tau}^\dagger(r) & \mathcal{L}_{\tau}^\dagger(r)\end{pmatrix}^T$ is used. Only one solution in shown in Eq.~\eqref{eq:35}. A second solution can be obtained by using time-reversal symmetry ($\Theta\Phi(r)$). The matching of  the wave functions in Eq.~\eqref{eq:35} at $r=0$ yields:
\begin{equation}\label{eq:36}
\begin{split}
     & \Delta_1(E+i\sqrt{\Delta_2^2-E^2}) = -\Delta_2(E-i\sqrt{\Delta_2^2-E^2})
     \\ & \to i[\Delta_1\sqrt{\Delta_2^2-E^2}-\Delta_2\sqrt{\Delta_1^2-E^2}] = -(\Delta_1+\Delta_2)E
     \\ & \to -2\Delta_1^2\Delta_2^2+2\Delta_1\Delta_2\sqrt{(\Delta_2^2-E^2)(\Delta_1^2-E^2)} = 2\Delta_1\Delta_2E^2
     \\ & \to \Delta_1^2\Delta_2^2 - (\Delta_1^2+\Delta_2^2)E^2 + E^4 = E^4 + 2\Delta_1\Delta_2E^2 + \Delta_1^2\Delta_2^2
     \\ & \to  0 = (\Delta_1+\Delta_2)^2E^2.
\end{split}
\end{equation}
Since $\Delta_1,\Delta_2>0$, $E=0$ according to Eq.~\eqref{eq:36}. Therefore, $\kappa_i=\Delta_i/\hbar v$ and the zero-energy bound states, $\Phi(r)$ and $\Theta\Phi(r)$, are:
\begin{equation}\label{eq:37}
\begin{split}
    \Phi(r) = & [\theta(r)e^{-\frac{\Delta_1}{\hbar v} r} + \theta(-r)e^{\frac{\Delta_2}{\hbar v} r}]\begin{pmatrix} i & 0 & 0 & 1 \end{pmatrix}^T,
    \\ \Theta\Phi(r) = & [\theta(r)e^{-\frac{\Delta_1}{\hbar v} r} + \theta(-r)e^{\frac{\Delta_2}{\hbar v} r}]\begin{pmatrix} 0 & i & 1 & 0 \end{pmatrix}^T,
\end{split}
\end{equation}
up to a normalization constant.

\par It is interesting to explore the robustness of the bound states under a wider range of conditions. Given their topological nature, we anticipate that these bound states will occur not only at sharp interfaces but also at smoother interfaces. To be concrete, we consider the following spatially dependent $\Delta_m$:
\begin{equation}
    \Delta_m(r) = 
    \begin{cases}
        \Delta\text{sgn}(r) & \text{for } |r|\geq r_0
        \\ \Delta\frac{r}{r_0} & \text{for } |r|< r_0
    \end{cases}.
\end{equation}

Here, $\Delta>0$ and $r_0>0$ are constants, with $r_0$ controlling the smoothness of the interface. Then, we use the ansatz $\Phi(r) = \theta(-r_0-r)\Phi_-(r)e^{\kappa r} +\frac{\text{sgn}(r_0-r)+\text{sgn}(r_0+r)}{2}\Phi_0(r)  + \theta(r-r_0)\Phi_+(r)e^{-\kappa r}$, where:
\begin{equation}\label{eq:35_new}
\begin{split}
    \Phi_-(r) & = \begin{pmatrix} \frac{-\Delta}{E+i\hbar v\kappa} & 0 & 0 & 1 \end{pmatrix}^T,
    \\ \Phi_0(r) & = \begin{pmatrix} \phi_1(r) & 0 & 0 & \phi_2(r) \end{pmatrix}^T,
    \\ \Phi_+(r) & = \begin{pmatrix} \frac{\Delta}{E-i\hbar v\kappa} & 0 & 0 & 1 \end{pmatrix}^T,
\end{split}
\end{equation}
where the functions $\phi_1(r),\phi_2(r)$ are to be determined. Equation~\eqref{eq:35_new} shows one of the solutions, while the second solution can be found through time-reversal symmetry ($\Theta\Phi(r)$). We can now obtain:
\begin{equation}\label{eq:36_new}
    \hbar v\kappa = \sqrt{\Delta^2-E^2} \text{ for }|r|\geq r_0,
\end{equation}
and:
\begin{equation}\label{eq:37_new}
    \begin{pmatrix} -i\hbar v\partial_r & \frac{\Delta}{r_0}r \\ \frac{\Delta}{r_0}r & i\hbar v\partial_r \end{pmatrix}\begin{pmatrix} \phi_1(r) \\ \phi_2(r) \end{pmatrix} = E\begin{pmatrix} \phi_1(r) \\ \phi_2(r) \end{pmatrix} \text{ for }|r|< r_0
\end{equation}
Application of the Gaussian ansatz, $\phi_i(r)=\alpha_ie^{-\frac{r^2}{2\xi_\text{loc}^2}}$ for $i=1,2$, to Eq.~\eqref{eq:37_new} yields:
\begin{equation}\label{eq:38_new}
\begin{split}
    \alpha_1(i\frac{\hbar v}{\xi_\text{loc}^2}r-E)+\frac{\Delta}{r_0}r\alpha_2 & = 0,
    \\ \frac{\Delta}{r_0}r\alpha_1 + \alpha_2(-i\frac{\hbar v}{\xi_\text{loc}^2}r-E) & = 0.
\end{split}
\end{equation}
According to Eq.~\eqref{eq:38_new}, $E=0$ and:
\begin{equation}\label{eq:39_new}
    \begin{pmatrix} i\frac{\hbar v}{\xi_\text{loc}^2} & \frac{\Delta}{r_0} \\ \frac{\Delta}{r_0} & -i\frac{\hbar v}{\xi_\text{loc}^2} \end{pmatrix}\begin{pmatrix} \alpha_1 \\ \alpha_2 \end{pmatrix} = 0.
\end{equation}
Matching solutions at $r=\pm r_0$ results in:
\begin{equation}
    \alpha_1 = i\alpha_2 = ie^{-\frac{\Delta}{2\hbar v}r_0},
\end{equation}
and: 
\begin{equation}\label{eq:40_new}
    \xi_\text{loc}^2 = \frac{\hbar v}{\Delta}r_0.
\end{equation}
Combining Eqs.~\eqref{eq:35_new} and \eqref{eq:40_new}, we obtain the zero-energy bound state wave function as:
\begin{equation}\label{eq:41_new}
\begin{split}
    \Phi(r) = & [\theta(-r_0-r)e^{\frac{\Delta}{\hbar v} r} + \frac{\text{sgn}(r_0-r)+\text{sgn}(r_0+r)}{2}e^{-\frac{\Delta }{2\hbar v r_0}(r_0^2+r^2)} + \theta(r-r_0)e^{-\frac{\Delta}{\hbar v} r}]\begin{pmatrix} i & 0 & 0 & 1 \end{pmatrix}^T,
    \\ \Theta\Phi(r) = & [\theta(-r_0-r)e^{\frac{\Delta}{\hbar v} r} + \frac{\text{sgn}(r_0-r)+\text{sgn}(r_0+r)}{2}e^{-\frac{\Delta }{2\hbar v r_0}(r_0^2+r^2)} + \theta(r-r_0)e^{-\frac{\Delta}{\hbar v} r}]\begin{pmatrix} 0 & i & 1 & 0 \end{pmatrix}^T,
\end{split}
\end{equation}
up to a normalization constant. Equation~\eqref{eq:41_new} indicates that the bound states possess a Gaussian profile for $|r|< r_0$ and evolve into exponential decay for $|r|\geq r_0$. For checking consistency, note that Eq.~\eqref{eq:41_new} reduces to Eq.~\eqref{eq:37} for $r_0 \to 0$. Then, the local density of states of the MKP, $\rho(r,E)$, derived from Eq.~\eqref{eq:41_new} is:
\begin{equation}
\begin{split}
    \rho(r,E) & = \text{tr}[\Phi(r)\Phi(r)^\dagger + \Phi_\Theta(r)\Phi_\Theta(r)^\dagger]\delta(E)
    \\ & =  \lim_{\Gamma\to0} \big[ \theta(-r_0-r)e^{\frac{2\Delta}{\hbar v} r} + \frac{\text{sgn}(r_0-r)+\text{sgn}(r_0+r)}{2}e^{-\frac{\Delta }{\hbar v r_0}(r_0^2+r^2)} + \theta(r-r_0)e^{-\frac{2\Delta}{\hbar v} r} \big]\frac{2\Gamma}{\pi(E^2+\Gamma^2)}, 
\end{split}
\end{equation}
up to a normalization constant, where $\Phi_\Theta(r)=\Theta\Phi(r)$ and $\Gamma$ is the spectral line broadening.

\section{S6. Fermion parity, soliton spin, and fermion number }
\par Since $[\varphi_\tau'(r),\frac{1}{\pi}\vartheta_\tau'(r')]=-i\frac{\text{sgn}(r-r')}{2}$ \cite{book:Giamarchi}, the fermion parity transforms $\vartheta_\tau'(r)$ into $\vartheta_\tau'(r)+\pi$:
\begin{equation}\label{eq:27}
\begin{split}
    e^{i\pi Q}\vartheta_\tau'(r)e^{-i\pi Q}  & = \vartheta_\tau'(r) + i\pi[Q,\vartheta_\tau'(r)]
    \\ & = \vartheta_\tau'(r) + i \int dr' [\partial_{r'}\varphi_\tau'(r'),\vartheta_\tau'(r)]
    \\ & = \vartheta_\tau'(r) + \pi \int dr' \partial_{r'}\frac{\text{sgn}(r-r')}{2}
    \\ & = \vartheta_\tau'(r) + \pi \int dr' \delta(r'-r)
    \\ & =  \vartheta_\tau'(r) + \pi.
\end{split}
\end{equation}
Here, we have denoted the fermion parity $(-1)^{N_F}$ as $e^{i\pi Q}$, where $Q$ is the fermion number of the fermionic soliton defined in the following section. To gain insight into the consequences of Eq.~\eqref{eq:27}, note that the bosonized Hamiltonian of Eq.~(9) of the main text before refermionization is:
\begin{equation}\label{eq:28}
    H_\tau(r) = \frac{\hbar v}{2\pi}\{[\partial_r\varphi_\tau'(r)]^2 + [\partial_r\vartheta_\tau'(r)]^2\} + \frac{2\Delta_m}{\pi a}\cos[2\vartheta_\tau'(r)].
\end{equation}
In view of Eq.~\eqref{eq:27}, $e^{i\pi Q}\cos[2\vartheta_\tau'(r)]e^{-i\pi Q}=\cos[2\vartheta_\tau'(r)+2\pi]=\cos[2\vartheta_\tau'(r)]$. Therefore, the Hamiltonian density in Eq.~\eqref{eq:28} respects the fermion parity. Further, when $\Delta_m>0$, $[\vartheta_\tau'(r)\mod2\pi]$ is pinned at $\frac{\pi}{2}$ and $\frac{3\pi}{2}$. But, when $\Delta_m<0$, $[\vartheta_\tau'(r)\mod2\pi]$ is pinned at $0$ and $\pi$, so that, in view of  Eq.~\eqref{eq:27}, the degenerate ground states are related by fermion parity, and the ground states respect the fermion parity, and $\ket{G_\pm}$ can be constructed as discussed in the main text.

\par Since $\vartheta_\tau'(r)$ is pinned at different values on different sides of the interface, the bound states described by Eq.~\eqref{eq:37} carry a $U(1)$ charge, $S_\tau'$, which equals the spin carried by the fermionic solitons:
\begin{equation}\label{eq:39}
    S_\tau' = \frac{1}{2\pi}\int dr \partial_r\vartheta_\tau'(r)
     = \frac{1}{2}\int dr \frac{1}{2\pi}\partial_r [\vartheta_\tau'(r)-\varphi_\tau'(r)] + [\vartheta_\tau'(r)+\varphi_\tau'(r)]
     = \frac{1}{2}\int dr [\rho_{\mathcal{R}_{\tau}}(r)-\rho_{\mathcal{L}_{\tau}}(r)]
     = \int dr \rho_{s'}(r).
\end{equation}
Here, $\rho_{\mathcal{R}_{\tau}(\mathcal{L}_{\tau})}(r)$ is the charge density of the right- (left-) moving soliton, and $\rho_{s'}(r)$ is the total spin density of the solitons. In the last line of Eq.~\eqref{eq:39}, we used the helical nature of the fermionic solitons. The total spin carried by the solitons equals the spin difference between the time-reversal sectors, which can be demonstrated by noticing that:
\begin{equation}\label{eq:40}
\begin{split}
    S_\tau' & = \frac{1}{2\pi}\int dr \partial_r\vartheta_\tau'(r)
     = \int dr \frac{\sqrt{2}}{2\pi}\partial_r\vartheta_\tau(r)
     = \int dr \frac{1}{2\pi}\partial_r [\vartheta_1(r)-\vartheta_2(r)]
    \\ &  = \frac{1}{2}\int dr \frac{1}{2\pi}\partial_r[\vartheta_1(r) - \varphi_1(r)] + \frac{1}{2\pi}\partial_r[\vartheta_1(r) + \varphi_1(r)] -\frac{1}{2\pi}\partial_r[\vartheta_2(r) - \varphi_2(r)] - \frac{1}{2\pi}\partial_r[\vartheta_2(r) + \varphi_2(r)]
    \\ & = \frac{1}{2}\int dr [\rho_{1R}(r)-\rho_{1L}(r)] - [\rho_{2R}(r) - \rho_{2L}(r) ]
     = \int dr [\rho_{1s}(r) - \rho_{2s}(r)]
     = S_1-S_2.
\end{split}
\end{equation}
Here, $\rho_{i\sigma}(r)$ is the charge density of the $\sigma=R, L$ mode in the $i$th sector, $\rho_{is}(r)$ is the spin density in the $i$th sector, and $S_i$ is the total spin of the $i$th sector. In the last line of Eq.~\eqref{eq:40}, we used the helical nature of the edge states.

\par We further show that $ S_\tau'\in\mathbb{Z}$ and $Q\in\mathbb{Z}$. Note first that the fermion number, $Q$, equals:
\begin{equation}\label{eq:42}
    Q = -\int dr\frac{1}{\pi}\partial_r\varphi_\tau'(r)
     = \int dr \frac{1}{2\pi}\partial_r[\vartheta_1(r) - \varphi_1(r)] - \frac{1}{2\pi}\partial_r[\vartheta_1(r) + \varphi_1(r)]
     = \int dr [\rho_{\mathcal{R}_{\tau}}(r)+\rho_{\mathcal{L}_{\tau}}(r)],
\end{equation}
which also equals half of the fermion number difference in different time-reversal sectors:
\begin{equation}\label{eq:42.5}
    Q = -\int dr\frac{1}{\pi}\partial_r\varphi_\tau'(r) 
     = -\int dr\frac{1}{\sqrt{2}\pi}\partial_r\varphi_\tau(r)
     = -\int dr\frac{1}{2\pi}\partial_r [\varphi_1(r)-\varphi_2(r)]
     = \int dr\frac{1}{2}[\rho_1(r)-\rho_2(r)]
     = \frac{1}{2}(N_1-N_2),
\end{equation}
where $\rho_i(r)$ is the total charge density of the $i$th sector and $N_i$ is the total charge in the $i$th sector. Since a uniform chemical potential is assumed such that $N_1=N_2$, $Q\neq0$ is characterized only by the transferred charge between the time-reversal sectors originating from some fluctuation. On the other hand, due to the interaction terms in Eqs.~\eqref{eq:S4} and \eqref{eq:S10}, the charge can transfer between modes with the same helicity on different time-reversal sectors in quantum fluctuations. Therefore,
\begin{equation}\label{eq:43}
    N_{1R} = N_0 + m, \quad N_{1L} = N_0 + n , \quad N_{2R} = N_0 - m, \quad N_{2L} = N_0 - n.
\end{equation}
Here, $N_{i\sigma}$ is the number of charges of the $\sigma=R, L$ mode in the $i$th sector, $N_0\in\mathbb{Z}$ is the number of charges before adding the interaction at equilibrium, and $n,m\in\mathbb{Z}$. This relationship involving the number of charges for different modes can be understood more easily if we write down the Hamiltonian densities for the mass terms in Eq.~\eqref{eq:S11} with fermionic operators through Eqs.~\eqref{eq:S4} and \eqref{eq:S10}:
\begin{equation}\label{eq:43.5}
\begin{split}
    H_\text{m}(r) = (V_{2k_0}-\tilde{V}_{2k_0})[:R^\dagger_1(r) R_2(r): :L^\dagger_1(r) L_2(r): + \text{H.c.}].
\end{split}
\end{equation}
Equation~\eqref{eq:43.5} implies that once a charge is annihilated in the right(left)-moving modes in one sector, the same number of charges is created in the right(left)-moving modes in another sector, leading to Eq.~\eqref{eq:43}. Then, after rewriting Eqs.~\eqref{eq:39} and \eqref{eq:42} as:
\begin{equation}\label{eq:44}
    S_\tau' = \frac{1}{2}[(N_{1R}-N_{1L})-(N_{2R}-N_{2L})] \text{ and } Q = \frac{1}{2}[(N_{1R}+N_{1L})-(N_{2R}+N_{2L})],
\end{equation}
we can obtain:
\begin{equation}\label{eq:45}
    S_\tau' = (m-n)\in\mathbb{Z} \text{ and } Q = (m+n)\in\mathbb{Z}.
\end{equation}
Furthermore, the fermion numbers of the right- (left-) moving mode of the soliton, $N_{\mathcal{R}_{\tau}}(N_{\mathcal{L}_{\tau}})$, are:
\begin{equation}\label{eq:46}
\begin{split}
    N_{\mathcal{R}_{\tau}} = & \int dr \frac{\partial_r[\vartheta_\tau'(r)-\varphi_\tau'(r)]}{2\pi} = \frac{Q}{2} + S_\tau' = \frac{3m-n}{2},
    \\ N_{\mathcal{L}_{\tau}} = & -\int dr \frac{\partial_r[\vartheta_\tau'(r)+\varphi_\tau'(r)]}{2\pi} = \frac{Q}{2} - S_\tau' = \frac{3n-m}{2}.
\end{split}
\end{equation}
Due to time-reversal symmetry, $m=n$, as seen from Eqs.~\eqref{eq:S4} and \eqref{eq:S10}, indicating that the fermionic solitons carry no net spin but net fermion number such that $N_{\mathcal{R}_{\tau}}=N_{\mathcal{L}_{\tau}}=n$ and $Q=2n$, of which a non-zero value holds only in some fluctuations when a uniform chemical potential is considered.

\section{S7. Physical meaning of the collective excitations and their time-reversal symmetry}
\par
\begin{figure}[ht]
  \centering
  \centering
    \includegraphics[width=0.75\linewidth]{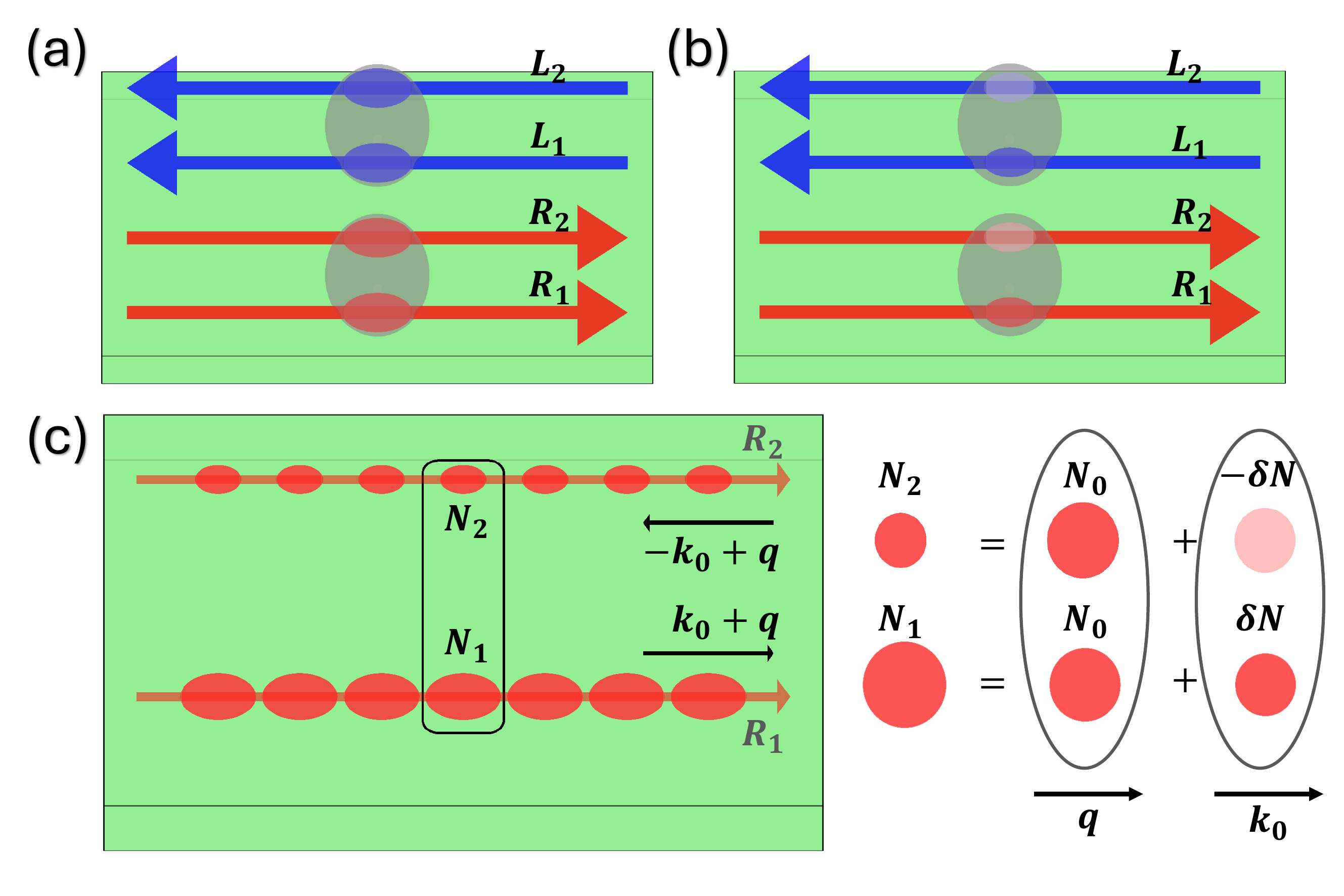}
  \caption{  A schematic diagram illustrating two types of collective excitations and how they are formed in interacting double helical edge states. (a) Helical electron-electron pairs in the charge sector of the bosonized Hamiltonian. (b) Helical electron-hole pairs in the pseudospin sector of the bosonized Hamiltonian. Filled (empty) spheres represent electrons (holes). Gray clouds around two spheres indicate the formation of electron-electron or electron-hole pairs. (c) Considering the right-moving mode of the first (second) time-reversal sector, $R_{1(2)}$, as an example, it is seen to carry $N_{1(2)}$ electrons (red spheres) with the plane-wave momentum $\pm k_0+q$ (black arrows). Motion of the particles in each time-reversal sector can be decomposed into a particle with fermion number $N_0=(N_1+N_2)/2$ and momentum $q$ in the charge sector, and another particle with fermion number $\pm\delta N=\pm(N_1-N_2)/2$ and momentum $k_0$ in the pseudospin sector. Intuitively, such a decomposition can be understood from momentum conservation: $N_1(k_0+q)+N_2(-k_0+q)=(N_1+N_2)q+(N_1-N_2)k_0$.}
  \label{fig:S3}
\end{figure}

 Here, we explore the collective excitations based on the picture in FIG.~\ref{fig:S3} in order to gain further insight into the nature of these excitations and their symmetry. Observe first that the fermionic fields in the refermionization of the pseudospin sector have the following analogous structures, 
\begin{equation}\label{eq:47}
\begin{split}
    \mathcal{R}_{\tau}(r) \propto & e^{i[\vartheta_\tau'(r)-\varphi_\tau'(r)]} \sim e^{i\sqrt{2}[\vartheta_\tau(r)-\varphi_\tau(r)]} \propto R^\dagger_2(r)R_1(r)
    \\ \mathcal{L}_{\tau}(r) \propto & e^{i[\vartheta_\tau'(r)+\varphi_\tau'(r)]} \sim e^{i\sqrt{2}[\vartheta_\tau(r)+\varphi_\tau(r)]} \propto L^\dagger_2(r)L_1(r).
\end{split}
\end{equation}
Accordingly, the right- (left-) moving solitons in the pseudospin sector can be viewed as a particle-hole pair on the right- (left-) moving modes, with the particle and hole coming from different time-reversal sectors, so that they carry a finite pseudospin and preserve the edge states' helical nature and Fermi velocity. This intuitively suggests that the effective model for the pseudospin sector must exhibit particle-hole symmetry, aligning with Eq.~(9) of the main text. 

Note that since the ``pseudospin'' arises from an antisymmetric combination of the time-reversal sectors, one can induce such pseudospin polarization by applying an out-of-plane electric field when time-reversal sectors are spatially separated, as in bilayer $\text{Bi}_\text{4}\text{Br}_\text{4}$ \cite{PhysRevB.109.155143} and coupled HgTe quantum wells \cite{PhysRevB.85.125309, PhysRevB.101.241302, PhysRevB.103.L201115, PhysRevB.93.235436, Ferreira2022, Krishtopenko2016}. As discussed already (Section S6. above), in edges with uniform chemical potential, the fermion number $Q$ of the soliton in the pseudospin sector is zero. However, due to the pairing in Eq.~(9) of the main text, $Q$ is not conserved and it is distinguished by the transfer of charges between the two time-reversal sectors originating from some fluctuation, see Eqs.~\eqref{eq:43} to \eqref{eq:46}. 

We can also refermionize the charge sector, which results in the following similarity in the bosonization language:
\begin{equation}\label{eq:49}
\begin{split}
    \mathcal{R}_c(r) \propto & e^{i[\vartheta_c'(r)-\varphi_c'(r)]} \sim R_2(r)R_1(r)
    \\ \mathcal{L}_c(r) \propto & e^{i[\vartheta_c'(r)+\varphi_c'(r)]} \sim L_2(r)L_1(r).
\end{split}
\end{equation}
The right(left)-moving solitons, $\mathcal{R}_c(r)$ and $\mathcal{L}_c(r)$, in the charge sector behave as particles on the right- (left-) moving modes on each time-reversal sector, and preserve the edge states' helical nature and carry finite charge. The solitons in the charge sectors can be shown to carry no spin but a net charge of $2N_0$, regardless of the presence of time-reversal symmetry. Furthermore, based on the picture in FIG.~\ref{fig:S3}, we can construct the soliton states in the charge and pseudospin sectors, $\psi_c(r)$ and $\psi_{\tau}(r)$, as:
\begin{equation}
\begin{split}
    \psi_{\tau}(r) & = \mathcal{R}_{\tau}(r)e^{ik_0r} + \mathcal{L}_{\tau}(r)e^{-ik_0r},
    \\ \psi_c(r) & = \mathcal{R}_c(r)e^{iqr} + \mathcal{L}_c(r)e^{-iqr},
\end{split}
\end{equation}
where $k_0\equiv(k_1+k_2)/2$ and $q\equiv(k_1-k_2)/2$. 

\par  From the physical picture outlined in FIG.~\ref{fig:S3}, spinful TRS must be maintained within each sector. This implies $\mathcal{R}_{\tau(c)} \to \mathcal{L}_{\tau(c)}$, $\mathcal{L}_{\tau(c)} \to -\mathcal{R}_{\tau(c)}$, and $i\to-i$ under TRS, resulting in $\varphi_{\tau(c)}'\to\varphi_{\tau(c)}'$ and $\vartheta_{\tau(c)}'\to-\vartheta_{\tau(c)}'+\pi$.  Therefore, under the time-reversal operation, $\ket{\frac{\pi}{2}}$ and $\ket{\frac{3\pi}{2}}$ transform into themselves while $\ket{0}$ and $\ket{\pi}$ transform into each other. $\ket{\frac{\pi}{2}}$, $\ket{\frac{3\pi}{2}}$, and $\frac{1}{\sqrt{2}}(\ket{0}+\ket{\pi})$ are thus TRS eigenstates with eigenvalue $1$ while $\frac{1}{\sqrt{2}}(\ket{0}-\ket{\pi})$ is the has eigenvalue $-1$. Here, $\ket{\vartheta_0}$ is the eigenstate of $\vartheta_\tau'$ such that $\vartheta_\tau'\ket{\vartheta_0}=\vartheta_0\ket{\vartheta_0}$. Hence, for $\Delta_m>0$, $\ket{G_\pm}$ in Eq.~(13) of the main text transform into themselves under TRS, preserving the TRS. For $\Delta_m<0$, $\ket{G_\pm}$ transform under TRS as $\ket{G_\pm}\to\pm\ket{G_\pm}$, indicating that $\ket{G_\pm}$ with $\Delta_m<0$ also preserves the TRS.

\section{S8. Extension to Hubbard interaction}
\par Instead of general density-density or current-current interactions, we turn now to consider the Hubbard interaction in the following form:
\begin{equation}
    H_\text{int}(r) = \frac{U}{2} \big[ \rho_\uparrow(r)\rho_\downarrow(r) + \rho_\downarrow(r)\rho_\uparrow(r) \big], 
\end{equation}
with a constant $U$. This corresponds to the special case where $V(r)=-\tilde{V}(r) = \frac{U}{2}\delta(r)$ for the potentials in Eq.~(3) of the main text. This interaction can be experimentally realized in cold-atom systems \cite{PhysRevA.92.053612, Gall2021, samland2024thermodynamicsdensityfluctuationsbilayer, Mitra2018}. The final bosonized Hamiltonian densities can be obtained using Eqs.~\eqref{eq:S11} to \eqref{eq:S12.5}:
\begin{align}
    H_{\text{LL}}(r) = & \frac{\hbar}{2\pi}\sum_{\mu=c,\tau}u_\mu[K_\mu^{-1}(\partial_r\varphi_\mu(r))^2 + K_\mu(\partial_r\vartheta_\mu(r))^2],
    \\ H_{\text{m}}(r) = & \frac{U}{2(\pi a)^2}\cos[2\sqrt{2}\vartheta_\tau(r)].
\end{align}
Here, the Luttinger liquid parameters and the renormalized velocities are:
\begin{equation}\label{eq:S76}
K_c = \sqrt{\frac{1-\frac{U}{\pi\hbar v}}{1+\frac{U}{\pi\hbar v}}},\, K_\tau =1 ,\, u_c=\frac{ v(1-\frac{U}{\pi\hbar v})}{K_c}, \, \text{and }u_\tau=\frac{v}{K_\tau}.
\end{equation}
The sign of the mass term is thus controlled by the sign of $U$, i. e. whether the Hubbard interaction is repulsive ($U>0$) or attractive ($U<0$). Interestingly, Eq.~\eqref{eq:S76} yields $K_c>1$ for $U<0$ and $K_c<1$ for $U>0$, so that the Hubbard interaction results in SC and $\pi$-SDW phases for $U<0$ and $U>0$, respectively.

\end{document}